\documentclass[paper]{JHEP3}
\pdfoutput=1
\usepackage{amsmath,amssymb,amsthm,amscd,graphicx}
\input epsf.sty

\newtheorem{theorem}{Theorem}[section]

\theoremstyle{definition}

\newtheorem{conjecture}[theorem]{Conjecture}

\newcommand{\CC}{{\cal C}}

\newcommand{\CG}{{\cal G}}

\newcommand{\CJ}{{\cal J}}

\newcommand{\CN}{{\cal N}}
\newcommand{\CO}{{\cal O}}

\newcommand{\CU}{{\cal U}}

\def\IZ{{\mathbb Z}}
\def\IR{{\mathbb R}}
\def\IC{{\mathbb C}}
\def\IP{{\mathbb P}}

\def\IS{{\mathbb S}}

\newcommand{\tr}{{\rm Tr}}
\newcommand{\re}{{\rm e}}
\newcommand{\ri}{{\rm i}}
\newcommand{\rd}{{\rm d}}
\newcommand{\rp}{{\rm p}}
\renewcommand{\d}{\partial}

\newcommand{\np}{{\mathrm{np}}}
\newcommand{\Tr}{\mathop{\rm Tr}\nolimits}
\newcommand{\Li}{\mathop{\rm Li}\nolimits}
\newcommand{\Ai}{\mathop{\rm Ai}\nolimits}
\newcommand{\Vol}{\mathrm{Vol}}
\newcommand{\wigner}{{\mathrm{W}}}
\newcommand{\GW}{{\mathrm{GW}}}
\newcommand{\ad}{\mathop{\rm ad}\nolimits}

\newcommand{\be}{\begin{equation}}
\newcommand{\ee}{\end{equation}}
\newcommand{\ba}{\begin{aligned}}
\newcommand{\ea}{\end{aligned}}
\newcommand{\ben}{\begin{eqnarray}\displaystyle}
\newcommand{\een}{\end{eqnarray}}

\newcommand{\sectiono}[1]{\section{#1}\setcounter{equation}{0}}

\newdimen\tableauside\tableauside=1.0ex
\newdimen\tableaurule\tableaurule=0.4pt
\newdimen\tableaustep
\def\phantomhrule#1{\hbox{\vbox to0pt{\hrule height\tableaurule width#1\vss}}}
\def\phantomvrule#1{\vbox{\hbox to0pt{\vrule width\tableaurule height#1\hss}}}
\def\sqr{\vbox{%
  \phantomhrule\tableaustep
  \hbox{\phantomvrule\tableaustep\kern\tableaustep\phantomvrule\tableaustep}%
  \hbox{\vbox{\phantomhrule\tableauside}\kern-\tableaurule}}}
\def\squares#1{\hbox{\count0=#1\noindent\loop\sqr
  \advance\count0 by-1 \ifnum\count0>0\repeat}}
\def\tableau#1{\vcenter{\offinterlineskip
  \tableaustep=\tableauside\advance\tableaustep by-\tableaurule
  \kern\normallineskip\hbox
    {\kern\normallineskip\vbox
      {\gettableau#1 0 }%
     \kern\normallineskip\kern\tableaurule}%
  \kern\normallineskip\kern\tableaurule}}
\def\gettableau#1{\ifnum#1=0\let\next=\null\else
\squares{#1}\let\next=\gettableau\fi\next}

\newcommand{\figref}[1]{Fig.~\protect\ref{#1}}
\title{ABJM theory as a Fermi gas}

\author{
Marcos Mari\~no and Pavel Putrov
\\
D\'epartement de Physique Th\'eorique et Section de Math\'ematiques,\\
Universit\'e de Gen\`eve, Gen\`eve, CH-1211 Switzerland\\
\\
\email{marcos.marino@unige.ch}, \quad
\email{pavel.putrov@unige.ch}
}

\abstract{The partition function on the three-sphere of many supersymmetric Chern--Simons--matter theories reduces, by localization, to a matrix model. 
We develop a new method to study these models in the M-theory limit, but at all orders in the $1/N$ expansion. The method is based on reformulating 
the matrix model as the partition function of an ideal Fermi gas with a non-trivial, one-particle quantum Hamiltonian. This new approach leads to a completely 
elementary derivation of the $N^{3/2}$ behavior for ABJM theory and $\CN=3$ quiver Chern--Simons--matter theories. In addition, 
the full series of $1/N$ corrections to the original matrix integral can be simply determined by a next-to-leading calculation in the 
WKB or semiclassical expansion of the quantum gas, and we show that,  for several quiver Chern--Simons--matter theories, it is given by an Airy function. 
This generalizes a recent result of Fuji, Hirano and Moriyama for ABJM theory. It turns out that the semiclassical expansion 
of the Fermi gas corresponds to a strong coupling expansion in type IIA theory, and it is dual to the genus expansion. This allows us to calculate explicitly 
non-perturbative effects due to D2-brane instantons in the AdS background. 
}

\begin{document}

\sectiono{Introduction}

One of the most interesting aspects of the AdS/CFT correspondence is that, in principle, one can use gauge theories to learn about elusive 
aspects of string theory and quantum gravity. For example, ABJM theory \cite{abjm}, as well as other supersymmetric 
Chern--Simons--matter (CSM) theories, are conjecturally dual to M-theory on 
spaces of the type AdS$_4 \times X_7$. Therefore, computations in the gauge theory side might give interesting insights on M-theory on these backgrounds.  

It was shown in \cite{kwy} that the partition function on the three-sphere of CSM theories with ${\cal N}\ge 3$ supersymmetry reduces, via localization, 
to a matrix model. This result was extended to theories with ${\cal N}\ge 2$ supersymmetry in \cite{jafferis,hama}. In the case of ABJM theory, the corresponding 
matrix model was solved for arbitrary 't Hooft coupling and at all orders in $1/N$ in \cite{dmp}, 
providing the first gauge theory derivation of the famous $N^{3/2}$ growth of the number 
of degrees of freedom of $N$ M2 branes \cite{kt}. 
In the last year, many results have been obtained for these matrix models, providing precision tests of the AdS$_4$/CFT$_3$ correspondence as well as beautiful 
field-theoretical results on supersymmetric 
CSM theories (see \cite{lectures} for a review and a list of references). 

Up to now, the ``stringy" side of these matrix models has been studied less intensively, but there have been already some interesting results in this direction for 
the ABJM theory. 
In \cite{dmp} it was found that the genus $g$ free energies of the matrix model contain very rich information about 
worldsheet instantons in type IIA theory (i.e. they include non-perturbative effects in $\alpha'$). In \cite{dmpnp}, 
by studying the large order behavior of the genus expansion, it was possible to identify as well 
non-perturbative corrections in the string coupling constant, conjecturally associated to D2-branes, or to 
membrane instantons in M-theory. Finally, building on the results of \cite{dmp,dmpnp}, it was shown in \cite{fhm} that, once worldsheet instantons are discarded, the full free energy of 
the ABJM matrix model is given by an Airy function. Schematically, we have 
\be
\label{Ai}
Z_{\rm ABJM} \propto {\rm Ai}\left[f(k)\left( N -g(k)\right) \right], 
\ee
where $k$ is the CS level (i.e. the inverse coupling), $f(k)\propto k^{1/3}$ is a function of $k$ determined by the large $N$ limit, and $g(k)$ is a function of $k$ which shifts $N$ (see (\ref{ZM}) below for a precise formula). This is a beautiful result which provides the all-orders expansion of the partition function in powers of the string length, and therefore resums the perturbative long-distance expansion for quantum superstrings.

This result raises an interesting possibility. The $N^{3/2}$ behavior of the free energy characterizes a large class 
of $\CN=3$ \cite{hklebanov,cmp} and $\CN=2$ \cite{ms,k1,jkps,ms2} Chern--Simons--matter theories. 
On the other hand, the result (\ref{Ai}) says that, for ABJM theory, 
this behavior is just the leading term in the logarithm of the Airy function. It is then natural to propose the following 

\begin{conjecture} \label{airycon} In parity-invariant supersymmetric CSM theories that display an $N^{3/2}$ growth in the number of degrees of freedom, 
the leading large $N$ limit 
and the $1/N$ corrections to the partition function on the three-sphere add up to an Airy function, i.e. we have schematically
\be
\label{conj}
Z_{\rm CSM} \propto {\rm Ai}\left[f(k_a) \left(N -g(k_a)\right)\right], 
\ee
where $k_a$ are the different CS levels involved in the theory, $f(k_a)$ is a function which is determined by the leading, large $N$ limit, and $g(k_a)$ is a shift which also depends on the details of the theory (but cannot be determined by the large $N$ limit alone). 
\end{conjecture}

If this conjecture is true, the corresponding matrix models display a {\it universal behavior} characterized by the Airy function. For theories which are not parity-invariant, we expect the Airy function to be the crucial ingredient of the answer. 
It is interesting to point out that the universal role of the Airy function in summing up $1/N$ corrections was already proposed in \cite{ovv}, in the different but related context of five-dimensional black holes made out of M2 branes in certain Calabi--Yau compactifications\footnote{We would like to thank C. Vafa for discussions on this.}. Unfortunately, the techniques used to derive (\ref{Ai}) for ABJM theory 
rely heavily on the calculation of $1/N$ corrections based on the holomorphic anomaly equation (see \cite{dmp,fhm} for details and references). For other models it is 
not clear how to generalize these techniques, even in cases (like the one studied in \cite{cmp}) where the planar resolvent is known explicitly.

All the results mentioned so far have been obtained in what we will call the 't Hooft expansion of the matrix model, which corresponds to the genus expansion in 
type IIA superstring theory. The 't Hooft expansion is the asymptotic expansion as $N$ goes to infinity and the 't Hooft parameter of the model, $\lambda$, is kept fixed at large $N$,
\be
\label{thooftl}
N \rightarrow \infty, \quad \lambda={N\over k}\, \,\,  \text{fixed}.
\ee
However, in order to make contact with M-theory, we have to consider the M-theory expansion, i.e. the asymptotic expansion as $N$ goes to infinity in which $k$ is fixed, 
\be
\label{Mtlimit}
N \rightarrow \infty, \quad k\, \, \text{fixed}. 
\ee
The 't Hooft expansion of the matrix model gives some information about the M-theory expansion. For example, the all-orders result (\ref{Ai}) presumably captures the all-orders expansion of the M-theory partition function in powers of the Planck length, at finite $k$. However, important contributions to the partition function in the M-theory expansion (like membrane instantons and Kaluza--Klein modes) are not directly captured in the 
't Hooft expansion, and in order to use the ABJM matrix model as a tool to explore M-theory, one should study the regime (\ref{Mtlimit}) directly. A step in this direction was 
taken in \cite{hklebanov}, who found a very simple method to extract the large $N$, fixed $k$ behavior of general CSM matrix models. However, in the approach of \cite{hklebanov} it is not obvious how to compute systematically $1/N$ corrections to the leading large $N$ behavior, not to speak about exponentially small corrections in $N$. 

It would then be very interesting to find a method to analyze the M-theory expansion of CSM theories, beyond the leading large $N$ contribution considered in 
\cite{hklebanov}. Such a method would allow us to address the above conjecture about the universal role of the Airy function in resumming the $1/N$ corrections, and 
eventually could give us information about M-theoretic features of the matrix models which are not manifest in the 't Hooft expansion. 

In this paper, we make a first step in this direction, and we propose a new method to analyze the matrix model of some Chern--Simons--matter theories which fulfills some of the expectations that we have just listed. The method consists of writing the partition function on the three-sphere as the partition function of an ideal Fermi gas with a non-trivial one-particle, quantum Hamiltonian. In this reformulation of the problem, the Chern--Simons level $k$ becomes the Planck's constant $\hbar$ of the quantum-mechanical problem. Since $k$ corresponds to the inverse string coupling, 
the semiclassical expansion of the Fermi gas is a {\it strong coupling expansion} in the type IIA theory large $N$ dual. 
As usual, the large $N$ limit is simply the thermodynamic limit of the gas. Our approach has the following features:

\vskip.3cm

1) The large $N$ limit of the free energy at finite $k$ is governed by the thermodynamic limit of the Fermi gas, which can be 
determined by a semiclassical calculation. In particular, 
we find a completely elementary derivation of the $N^{3/2}$ behavior of the free energy of ABJM theory, including the correct coefficient. 
This can be extended in a straightforward way to $\CN=3$ necklace quivers, and the result for the free energy 
is in full agreement with the calculation in \cite{herzog,herzog2} based on the matrix model analysis of \cite{hklebanov}. Interestingly, 
the relevant Fermi surface describing the thermodynamic limit of the Fermi gas is a two-dimensional polytope which characterizes the geometry of the 
dual tri-Sasakian spaces. In the case of ABJM theory, this Fermi surface is also a real version of the tropical curve obtained in \cite{cmp}. 

\vskip.3cm
2) The full $1/N$ expansion of the free energy of the matrix model is determined by the first quantum correction to the semiclassical limit. In this way we 
reproduce the result (\ref{Ai}) for ABJM theory at finite $k$, using again elementary methods in quantum Statistical Mechanics. Moreover, we prove our conjecture (\ref{airycon}) for 
a large class of $\CN\ge 3$ supersymmetric CSM theories. 

\vskip.3cm
3)  One can also compute exponentially suppressed effects at large $N$ which are clearly M-theoretic. In particular, we find a systematic method to determine the contribution of membrane instantons. These non-perturbative effects receive however corrections at all orders in the $\hbar$ expansion, and so far we can only determine them order by order in $k$ (but to all orders in the membrane winding). The strength of these corrections (i.e. the minimal membrane action) agrees with the instanton analysis of \cite{dmpnp}.

\vskip.3cm

It follows from the last point above that, at the level of non-perturbative corrections, our method does not fully capture the M-theory expansion, since so far we are only able to determine these corrections in an expansion in $k$ around $k=0$. In order to make contact with the true M-theory expansion one should resum the resulting series. In spite of this limitation, the Fermi gas picture gives a concrete computational method to address non-perturbative effects in these superstring theory backgrounds. In fact, the semiclassical expansion of the Fermi gas is 
dual to the conventional genus expansion captured by the 't Hooft expansion of the matrix model. 
For example, in the 't Hooft expansion, worldsheet instantons appear as exponential corrections in the 't Hooft coupling, order by order in the $g_s$ expansion, 
while membrane corrections appear as large $N$ instantons, of order $\exp(-1/g_s)$. In the Fermi gas approach developed in this paper, membrane instantons appear as exponential corrections in the chemical potential of the gas, order by order in the $\hbar$ expansion, while worldsheet instantons appear as quantum-mechanical instanton effects of order $\exp(-1/\hbar)$.
 
Finally, we would like to point out that the use of Fermi gas techniques in the analysis of matrix models goes back to the solution of matrix quantum mechanics in \cite{bipz}, and should be familiar from the study of the $c=1$ string (see for example \cite{klebanov}). The idea of studying the matrix integral partition function in the grand-canonical ensemble appeared in \cite{bk}, and was developed in detail in \cite{kostov,kkn,hkk}. In particular, the semiclassical limit of the Fermi gas was already used in Appendix A of \cite{kkn}. The systematic application of semiclassical techniques of many-body physics in the study of these matrix integrals, which we develop in this paper, seems however to be new. 

The paper is organized as follows. In section 2 we review previous results on the $1/N$ expansion of the ABJM matrix model in the 't Hooft expansion, 
focusing on \cite{dmp,dmpnp,fhm}. In section 3 we show that the matrix integral of a general class of $\CN\ge 3$ CSM theories (necklace quivers with fundamental matter) 
can be written as the partition function of an ideal Fermi gas with a non-trivial one-particle 
Hamiltonian. In sections 4 and 5 we present the tools to analyze the Fermi gas, and we illustrate them in ABJM theory. More precisely, in section 4 we study the Fermi gas in the thermodynamic limit, by passing to the grand canonical ensemble. This makes it possible to derive the leading $N^{3/2}$ behavior of the free energy of ABJM theory, by using elementary tools in Statistical Mechanics. We also compute exponentially suppressed corrections to the grand canonical potential, which are interpreted as membrane instantons in M-theory.  
In section 5 we study the quantum corrections to the grand canonical potential. We show that, up to non-perturbative terms, a next-to-leading WKB 
calculation is enough to determine the full $1/N$ expansion of the canonical free energy. This provides a simple derivation of the Airy function resummation 
of \cite{fhm}. In section 6 we extend our techniques to more general CSM theories, including necklace quivers and theories with fundamental matter. We show that, when 
the free energy on the three-sphere is real, the $1/N$ expansion at fixed $k$ gets resummed by 
an Airy function, thus proving conjecture \ref{airycon} for this family of examples. We also consider the 
``massive" theory of \cite{gt}, where a different $N^{5/3}$ scaling has been found for the free energy, and we rederive it with our techniques. Finally, in section 8 we 
conclude with some prospects for future work. In an Appendix we collect some results for the grand canonical potential of ABJM theory at order $\CO(\hbar^4)$. 

As this paper was being prepared for submission, the paper \cite{okuyama} appear which also considers ABJM theory in the grand canonical ensemble. 

\sectiono{The ABJM matrix model in the 't Hooft expansion}

\subsection{$1/N$ expansion and non-perturbative effects}

The matrix integral describing the partition function of ABJM theory on $\IS^3$ is given by \cite{kwy}
\be
\label{abjmmatrix}
\ba
&Z_{\rm ABJM}(N)\\
&={1\over N!^2} \int {\rd ^N \mu \over (2\pi)^N} {\rd ^N \nu \over (2\pi)^N} {\prod_{i<j} \left[ 2 \sinh \left( {\mu_i -\mu_j \over 2} \right)\right]^2
  \left[ 2 \sinh \left( {\nu_i -\nu_j \over 2} \right)\right]^2 \over \prod_{i,j} \left[ 2 \cosh \left( {\mu_i -\nu_j \over 2} \right)\right]^2 } 
  \exp \left[ {\ri k \over 4 \pi} \sum_{i=1}^N (\mu_i^2 -\nu_i^2) \right].
  \ea
  \ee
This matrix integral can be solved in the 't Hooft expansion (\ref{thooftl}) by using techniques of matrix model theory and topological string theory \cite{mp,dmp}. 
In particular, one can obtain explicit formulae for the genus $g$ free energies appearing in the $1/N$ expansion
\be
F(\lambda,g_s)=\sum_{g=0}^{\infty} g_s^{2g-2} F_g(\lambda),
\ee
where the 't Hooft coupling $\lambda$ is defined in (\ref{thooftl}), and 
\be
g_s ={2 \pi \ri \over k}.
\ee
The genus $g$ free energies $F_g(\lambda)$ obtained in this way are exact interpolating functions, and they can be studied in various regimes of the 
't Hooft coupling. When $\lambda\rightarrow 0$ they reproduce the perturbation theory of the matrix model around the Gaussian point. 
They can be also studied in the strong coupling regime $\lambda \rightarrow \infty$, where one can make contact with the AdS dual. In this regime it is more convenient to use the shifted variable 
\be
\label{hatl}
\hat \lambda =\lambda -{1\over 24}.
\ee
As explained in \cite{dmp}, this shift is expected from type IIA and M-theory arguments \cite{bh,ahho}. It turns out that, when expanded at strong coupling, the genus 
$g$ free energies have the structure
\be
F_g(\hat \lambda)=F_g^{\rm p}( \hat \lambda) + F_g^{\rm np}( \hat \lambda). 
\ee
The first term represents the perturbative contribution in $\alpha'$, while the second term is non-perturbative in $\alpha'$, 
\be
\label{winst}
F_g^{\rm np}(\hat  \lambda) \sim \CO\left(\re^{-2 \pi {\sqrt{2 \hat \lambda}}}\right)
\ee
and it was interpreted in \cite{dmp} as the contribution of worldsheet instantons in the type IIA dual. For $F_{0,1}$, the perturbative part is of the form, 
\be
\ba
F_0^{\rm p}(\hat  \lambda)&={4 \pi^3 {\sqrt{2}} \over 3} \hat \lambda^{3/2}, \\
F_1^{\rm p}(\hat  \lambda) &={\pi\over 6} {\sqrt{2 \hat \lambda}}  -{1\over 2} \log\left[ 2 {\sqrt{2 \hat \lambda}} \right], 
\ea
\ee
while for $g\ge 2$ one has
\be
\label{fgl}
F_g^{\rm p}(\hat  \lambda)= f_g\left( {1\over {\sqrt{\hat \lambda}}} \right),
\ee
where 
\be
f_g(x)=\sum_{j=0}^g c_j^{(g)}x^{2g-3+j}
\ee
is a polynomial. 

Besides the non-perturbative effects in $\alpha'$, one can use the connection between the large-order behavior of perturbation theory and instantons to deduce the 
structure of non-perturbative effects in the string coupling constant. In \cite{dmpnp} a detailed analysis showed that these effects would have the form 
\be
\label{membranea}
\exp\left( -k \pi {\sqrt{2 \lambda}} \right) 
\ee
at large $\lambda$. These were interpreted as D2-branes wrapped around generalized Lagrangian cycles of the target geometry. We will refer to these non-perturbative effects as membrane instanton effects, since they can be interpreted as M2 instantons in M-theory \cite{bbs} but they are invisible in ordinary string perturbation theory. 

\subsection{The partition function as an Airy function}

It was shown in \cite{fhm} that the genus expansion of the perturbative free energies can be resummed. In order to do that, one has to use the variable \cite{dmpnp}
\be
\label{extrashift}
\lambda_{\rm ren}=\lambda-{1\over 24} -{1\over 3k^2}
\ee
rather than (\ref{hatl}). If we define the perturbative partition function as 
\be
Z_{\rm ABJM}^{\rp}= \exp \left[ \sum_{g=0}^{\infty} F_g^{\rm p}(\hat \lambda) g_s^{2g-2}\right] 
\ee
then 
\be
Z_{\rm ABJM}^{\rp}\propto {\rm Ai}\left[ \left( {\pi^2 k^4 \over  2} \right)^{1/3} \lambda_{\rm ren}\right], 
\ee
where ${\rm Ai}$ is the Airy function. This can be also written in terms of $N$ as
\be
\label{ZM}
Z_{\rm ABJM}^{\rp}\propto {\rm Ai}\left[ C^{-1/3} \left( N-{k\over 24} -{1\over 3k} \right) \right],
\ee
where
\be
\label{Ck}
C={2 \over \pi^2 k}. 
\ee
As noticed in \cite{dmpnp}, the expansion resummed in (\ref{ZM}) makes perfect sense for finite $k$. Therefore, even if (\ref{ZM}) was obtained from a calculation in the 
't Hooft expansion, it should be part of the M-theory answer. Indeed, one of our goals in this paper is to verify this by computing $Z_{\rm ABJM}$ directly in the M-theory expansion. 

The Airy function appearing in (\ref{ZM}) gives an exact resummation of the long-distance expansion in M-theory. To see this, one has to use the dictionary 
relating gauge theory quantities to gravity quantities. 
In particular, one has to take into account the anomalous shifts relating the rank of the gauge group $N$ to the Maxwell charge $Q$, which in turn determines the compactification 
radius $L$ \cite{bh, ahho}. The relation is 
\be
Q=N-{1\over 24}\left( k-{1\over k}\right).
\ee
The charge $Q$ determines the compactification radius in M-theory according to 
\be
\left( {L \over \ell_p}\right)^6= 32 \pi^2 Q k, 
\ee
where $\ell_p$ is the Planck length. The shift (\ref{extrashift}) was interpreted in \cite{dmpnp} as a renormalization of the expansion parameter $\ell_p/L$, since 
it means that the natural variable is 
\be
\label{renc}
{\widehat \ell_p\over L} ={\ell_p / L\over \left[ 1- 12 \pi^2 \left( \ell_p/L\right)^6 \right] ^{1/6}},
\ee
and then the argument of the Airy function (\ref{ZM}) is given by 
\be
\left( 256 \, k \, \pi^2 \right)^{-2/3} \left({ L \over \widehat \ell_p} \right)^6.
\ee

The $1/N$ expansion of the ABJM matrix model was derived in \cite{dmp} by using the holomorphic anomaly equations \cite{bcov} of topological string theory. 
The result (\ref{ZM}) was obtained in \cite{fhm} by looking at the recursive structure of these equations. There is however a much simpler method to 
obtain (\ref{ZM}) which exploits the wavefunction behavior of the topological string partition function\footnote{We would like to thank C. Vafa for reminding us this.}. 
Our derivation of (\ref{ZM}) in this paper does not depend at all on ideas from topological string theory, but since it is formally very similar, we will now present 
this simpler argument. We will rely on results and notations of \cite{dmp}. Readers who are not familiar with topological string theory can skip the rest of this section 
and proceed to the next one.  

As shown in \cite{wittenwave}, it follows from the holomorphic anomaly equations that the topological string partition function is a wavefunction on moduli space. In particular, its transformation from one symplectic frame to the other is given by a Fourier transform. This property was spelled out in detail and exploited in \cite{abk}. 
The main result is summarized as follows. Let 
\be
\label{symplec}
\Gamma=\begin{pmatrix} \alpha & \beta \\ \gamma & \delta \end{pmatrix} \in {\rm SL}(2, \IZ)
\ee
be a symplectic transformation relating two different frames (we assume for simplicity that there is a single modulus in the problem). 
This means that the periods $(\partial_a F_0, a)$ transform as 
\be
\begin{pmatrix} \partial_{a^\Gamma} F^{\Gamma}_0  \\  a^{\Gamma} \end{pmatrix}=\Gamma \begin{pmatrix} \partial_{a} F_0  \\ a \end{pmatrix}.
\ee
Then, the full topological string partition function 
\be
Z(a) =\exp\left[ \sum_{g=0}^{\infty}  F_g(a) g_s^{2g-2} \right]
\ee
transforms as
\be
\label{zfou}
 Z^{\Gamma} (a^{\Gamma})=\int \rd a \, \re^{-S(a, a^{\Gamma})/g_s^2 } Z(a),
\ee
where
\be
\label{sfou}
S(a, a^{\Gamma})= -{1\over 2} \delta \gamma^{-1} a^2 + \gamma^{-1} a a^{\Gamma} -{1\over 2} \alpha \gamma^{-1} \left(a^{\Gamma}\right)^2. 
\ee

In the context of ABJM theory, as explained in detail in \cite{mp,dmp}, the relevant 
quantities correspond to topological string computations in the so-called orbifold frame, where the natural periods are $\lambda$ (the 
't Hooft coupling of the gauge theory) and the derivative $\partial_\lambda F_0$. On the other hand, the most familiar frame in topological string theory is the large radius or Gromov--Witten 
frame, where the natural periods are $T$ (the K\"ahler modulus) and the derivative $\partial_T F_0^{\rm GW}$. The genus $g$ free energies in the large radius frame are given by the 
standard formulae, 
\be
\label{LRfree}
\ba
 F^\GW_0&=\frac{T^3}{6}+\sum_{k>0}N_{0,k} \re^{-kT},\\
 F^\GW_1&=\frac{T}{12}+\sum_{k>0}N_{1,k} \re^{-kT},\\
 F^\GW_{g>1}&=\sum_{k>0}N_{g,k} \re^{-kT},
\ea
\ee
where $N_{g,k}$ are Gromov--Witten invariants in the local $\IP^1 \times \IP^1$ geometry (there is no constant term contribution at higher genus). 
The fact that the total free energy is at most cubic in $T$, up to exponentially small 
corrections, is a well known fact in topological string theory. 

In \cite{dmp}, the periods in the orbifold frame were written in terms of periods in the large radius frame in order to perform analytic continuations to strong coupling. By general principles, 
this relation must be a symplectic transformation like (\ref{symplec}). In fact, it is easy to see that the results of \cite{dmp} relating the periods can be written as the following symplectic transformation:
\begin{equation}
 \left(\begin{array}{c}
     \partial_{\tilde {\lambda}}\widetilde F_0 \\ \tilde{\lambda}
       \end{array}\right)=
\left(\begin{array}{cc}
 0 & 1 \\
 -1 & 2
\end{array}\right)
\left(\begin{array}{c}
    \partial_{\widetilde{T}} \widetilde{F}^{\rm GW}_0 \\  \widetilde{T}
       \end{array}\right)
\end{equation}  
where 
\be
\tilde{\lambda}={4\pi^2 \over c} \lambda, \quad \widetilde{T}={\pi \ri \over 2c} T, \qquad c^2=2\pi \ri,
\ee
and
\be
\label{shift}
\ba
\widetilde F_0&=F_0-\pi^3\ri\lambda,\\
 \widetilde{F}^\GW_{g}&=(-4)^{g-1}\left(F^\GW_{g}-\delta_{g,0}\frac{\pi^2T}{3}\right).
 \ea
\end{equation}
Then, according to (\ref{zfou}), (\ref{sfou}), the total partition functions are related by the following formula:
\be
 \exp\left[F(\lambda)-\pi^3\ri\lambda/g_s^2\right] \propto \int \rd\widetilde{T}
\exp\left[-\widetilde{T}^2/g_s^2+\widetilde{T}\tilde{\lambda}/g_s^2+\widetilde{F}^{\rm GW}(\widetilde{T})\right].
\ee
Notice that, up to nonperturbative terms in $T$, this is the integral of the exponential of a cubic polynomial, therefore 
we will indeed get an Airy function. Let us introduce the new variable $\mu$ through 
\be
\label{Tmu}
T={4 \mu \over k}-\pi \ri. 
\ee
Then, one finds the expression
\be
\label{express}
\ba
 \exp F(\lambda)& \propto 
\int \rd \mu
\exp\left\{\frac{2\mu^3}{3 k\pi^2}-\mu N+\frac{k}{24}\,\mu+\frac{1}{3k}\,\mu+\CO\left(\re^{-\frac{4\mu}{k}}\right)\right\}\\ 
&\propto \Ai\left[C^{-1/3}(N-B)\right]\left(1+{\CO}(\re^{-2 \pi {\sqrt{2 \lambda}}})\right),
\ea
\ee
where we used the following integral representation of the Airy function, 
\be
{\rm Ai}(z)={1\over 2 \pi \ri} \int_{\CC} \rd t\, \exp\left( {t^3 \over 3} -z t\right), 
\ee
and $\CC$ is a contour in the complex plane from $\re^{-\ri \pi/3}\infty$ to $\re^{\ri \pi/3}\infty$. In (\ref{express}), $C$ is given in (\ref{Ck}) and 
\be
B={k\over 24}+{1 \over 3k}. 
\ee
The result of (\ref{express}) is of course the expression obtained in (\ref{ZM}). Notice that the first term in the shift $B$ comes from $F_0^{\rm GW}$ in (\ref{shift}), while the 
second term is due to the first, perturbative term in $F_1^{\rm GW}$. The exponentially small corrections in $N$ in (\ref{express}), which are due to  
the worldsheet instantons at large radius of the topological string, become, after Fourier transform, the 
worldsheet instantons (\ref{winst}) of the type IIA superstring. 

This derivation is nice, but it seems difficult to generalize it in its current form to other 
Chern--Simons--matter theories, and prove in this way the conjecture (\ref{conj}) for other cases. In this paper we will 
find a completely different approach to the derivation of the Airy function which turns out to formally equivalent to the 
one based on topological string theory. However, this approach 
can be extended to many $\CN\ge3$ CSM theories and makes it possible to verify the conjecture \ref{airycon} for many of them.

\sectiono{Chern--Simons--matter theories as Fermi gases}

\subsection{ABJM theory as a Fermi gas}
\label{rewriting}

Our Fermi gas approach is based on the following observation. The interaction term between 
the eigenvalues in (\ref{abjmmatrix}) can be written in a different way by using the Cauchy identity:
 \be
 \label{cauchy}
 \ba
  {\prod_{i<j}  \left[ 2 \sinh \left( {\mu_i -\mu_j \over 2}  \right)\right]
\left[ 2 \sinh \left( {\nu_i -\nu_j   \over 2} \right) \right] \over \prod_{i,j} 2 \cosh \left( {\mu_i -\nu_j \over 2} \right)}  
 & ={\rm det}_{ij} \, {1\over 2 \cosh\left( {\mu_i - \nu_j \over 2} \right)}\\
 &=\sum_{\sigma \in S_N} (-1)^{\epsilon(\sigma)} \prod_i {1\over 2 \cosh\left( {\mu_i - \nu_{\sigma(i)} \over 2} \right)}.
 \ea
  \ee
  In this equation, $S_N$ is the permutation group of $N$ elements, and $\epsilon(\sigma)$ is the signature of the permutation $\sigma$. 
  This identity has been used in other matrix models in \cite{bk,kostov,kkn} in order to study them in the grand canonical ensemble, as we will do here. 
In the context of ABJM theory, it was used in \cite{kwytwo} in order to prove the equivalence of (\ref{abjmmatrix}) and the matrix integral for $\CN=8$ 
super Yang--Mills theory in three dimensions, when $k=1$. The manipulations in \cite{kwytwo} can be easily generalized to arbitrary $k$, 
and one obtains the following expression for the ABJM matrix model, 
\be
\label{fgasform}
Z(N)={1 \over N!} \sum_{\sigma  \in S_N} (-1)^{\epsilon(\sigma)}  \int  {\rd ^N x \over (2 \pi k)^N} {1\over  \prod_{i} 2 \cosh\left(  {x_i  \over 2}  \right)
2 \cosh\left( {x_i - x_{\sigma(i)} \over 2 k} \right)}.
\ee
We will derive this expression below with a different technique, which can be used for more general Chern--Simons--matter theories. The main property of (\ref{fgasform}) is 
that it makes contact with the standard formalism to study partition functions of ideal Fermi gases. Indeed, let us introduce the function 
\be
\label{densitymat}
\rho(x_1, x_2)={1\over 2 \pi k} {1\over \left( 2 \cosh  {x_1 \over 2}  \right)^{1/2} }  {1\over \left( 2 \cosh {x_2  \over 2} \right)^{1/2} } {1\over 
2 \cosh\left( {x_1 - x_2\over 2 k} \right)}.
\ee
If we interpret it as a one-particle density matrix in the position representation 
\be
\rho(x_1, x_2) =\langle x_1 | \hat \rho |x_2 \rangle,
\ee
the matrix integral (\ref{abjmmatrix}) can be written as the partition function of an ideal Fermi gas with $N$ particles
\be
\label{zabjm}
Z(N)={1 \over N!} \sum_{\sigma  \in S_N} (-1)^{\epsilon(\sigma)}  \int  \rd ^N x \prod_i \rho(x_i, x_{\sigma(i)}).
\ee
It is well-known that the sum over permutations appearing in the canonical free energy of an ideal quantum gas can be written as a sum over conjugacy classes of the permutation group (see for example \cite{feynman}). A conjugacy class is specified by a set of integers $\{ m_{\ell} \}$ satisfying
\be
\label{Ncons}
\sum_\ell \ell m_\ell=N.
\ee
Let us define
\be
\label{zell}
Z_\ell = \int \rd x_1 \cdots \rd x_\ell   \, \rho(x_1, x_2)\rho(x_2, x_3)\cdots \rho(x_{\ell-1}, x_\ell) \rho(x_{\ell}, x_1).
\ee
Then, the partition function is given by, 
\be
\label{conjclasses}
Z(N) =\sum_{\{ m_\ell \}} {}^{'}\prod_\ell   {\eta^{(\ell-1)m_\ell } Z_\ell^{m_\ell} \over m_\ell! \ell^{m_\ell}}
\ee
where the $ {}^{'}$ means that we only sum over the integers satisfying the constraint (\ref{Ncons}). 

Due to the constrained sum, the canonical partition function is not easy to handle for large $N$. As usual, the remedy is to consider the grand partition function 
\be
\label{grand}
\Xi=1+\sum_{N=1}^\infty Z(N) z^N, 
\ee
where 
\be
z=\re^{\mu}
\ee
plays the r\^ole of the fugacity and $\mu$ is the chemical potential. The grand-canonical potential is 
\be
J(\mu) =\log \Xi. 
\ee
Notice that this potential (like the free energy) has the opposite sign to the usual conventions in Statistical Mechanics. A standard argument (presented for example in 
\cite{feynman}) tells us that the sum over conjugacy classes in (\ref{conjclasses}) can be written as
\be
\label{jsum}
J(\mu)=-\sum_{\ell \ge 1} Z_\ell {(-z)^\ell \over \ell}.
\ee
The canonical partition function is recovered from the grand-canonical potential as
\be
\label{exactinverse}
Z(N) =\oint {\rd z \over 2 \pi \ri } {\Xi \over z^{N+1}}.
\ee
At large $N$, this integral can be computed by applying the saddle-point method to
\be
\label{muint}
Z(N) ={1\over 2 \pi \ri} \int \rd \mu \, \exp\left[J(\mu) - \mu N\right].
\ee
The saddle point occurs at
\be
N =  {\partial J \over \partial \mu}=-\sum_{\ell \ge 1} Z_\ell (-z)^\ell,
\ee
and defines a function $\mu_*(N)$. The free energy is given, at leading order as $N \rightarrow \infty$, by
\be
\label{freethermo}
F(N) =J(\mu_*) - \mu_* N.
\ee
However, it is possible to compute the $1/N$ corrections to this relation by simply computing the corrections to the full integral in 
(\ref{muint}). This is what we will eventually do. Notice the similarity between the traditional inverse transform (\ref{muint}) and the Fourier transform (\ref{express}) in 
topological string theory.
\begin{figure}
\center
\includegraphics[height=4cm]{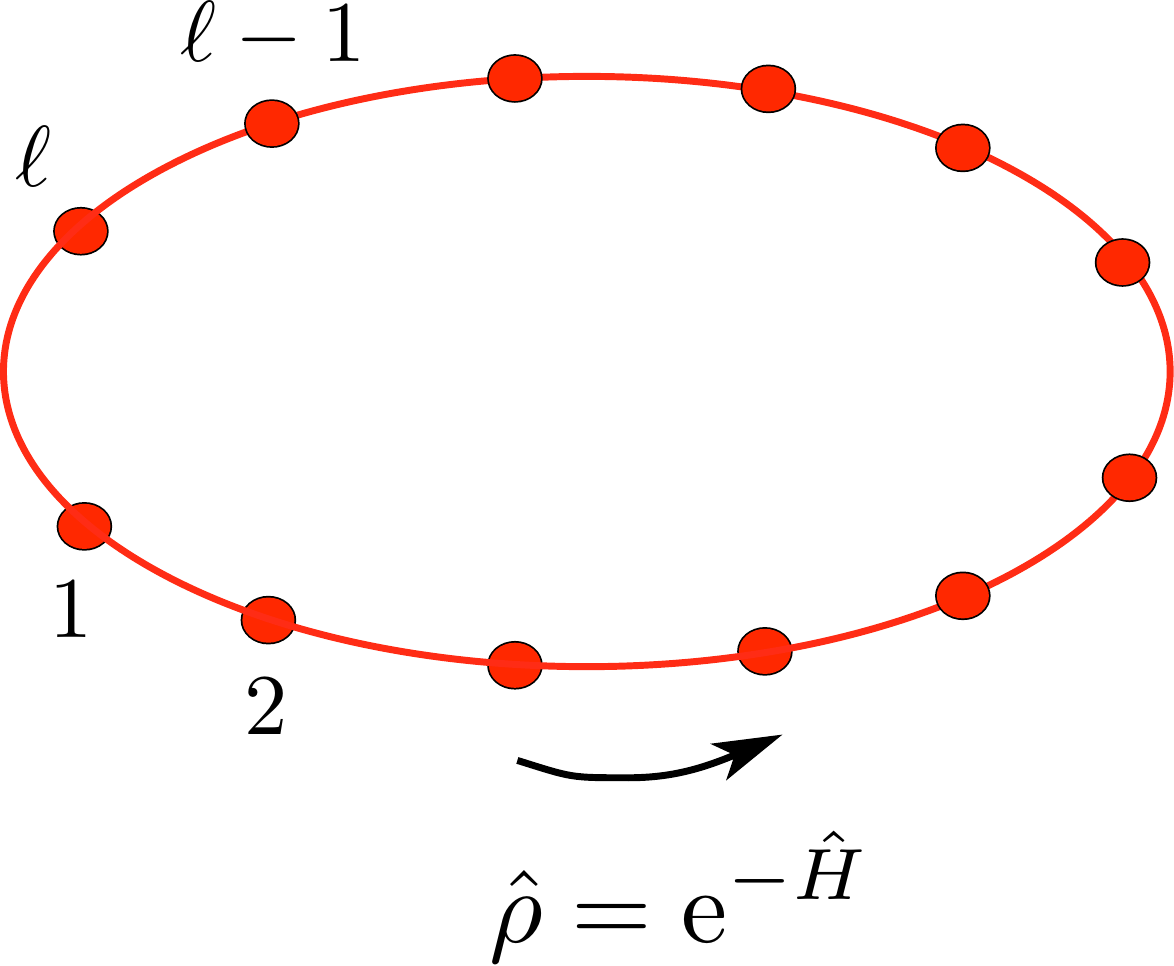} 
\caption{A one-dimensional 
periodic lattice with $\ell$ sites. The transfer matrix $\hat\rho$ can be regarded as the quantum propagator for a single particle in Euclidean, discretized time with Hamiltonian $\hat H$.}
\label{lattice}
\end{figure}

We have then shown that the original ABJM matrix integral can be computed as the canonical partition function of a system of $N$ non-interacting fermions, where the one-particle density matrix is given by (\ref{densitymat}). We just have to solve the corresponding one-body problem in order to compute the relevant thermodynamic quantities of the system. Equivalently, one should compute the quantity $Z_\ell$ introduced in (\ref{zell}). This quantity can be regarded as the partition function of a classical lattice gas with $\ell$ particles in a periodic lattice with nearest-neighbour interactions, as shown in \figref{lattice}. The density matrix $\rho$ plays the r\^ole of the 
 classical transfer matrix of the system (see for example chapter 12 of \cite{parisi}). It defines a symmetric kernel 
 \be
\langle x| \hat  \rho | \phi\rangle=\int \rd x' \, \rho(x, x') \phi (x'), 
 \ee
 so that 
 \be
 \label{zellrho}
 Z_\ell =\tr\, \hat \rho^\ell. 
 \ee
 It is easy to see that this kernel is a non-negative Hilbert--Schmidt operator, therefore it has a discrete, positive spectrum 
 \be
 \label{eigenv}
\hat \rho|\phi_n \rangle=\lambda_n |\phi_n\rangle, \qquad n=0, 1, \cdots,
 \ee
 where $|\phi_n \rangle$ are orthonormal eigenfunctions and we assume that
\be
\lambda_0 \ge \lambda_1 \ge \lambda_2 \ge \cdots. 
\ee
We can then write the density matrix as 
 \be
 \hat \rho=\sum_{n\ge0} \lambda_n |\phi_n\rangle \langle  \phi_n|.
\ee
In terms of these eigenvalues we have, 
\be 
Z_\ell =\sum_{n \ge0} \lambda_n^{\ell}. 
\ee
When $\ell$ is large, this sum is dominated by the largest eigenvalue $\lambda_0$, 
\be
Z_\ell \approx \lambda_0^{\ell}, \qquad \ell \gg 1. 
\ee
It also follows from this representation that the grand-canonical partition function is given by a Fredholm determinant, 
\be
\label{fred-det}
\Xi={\rm det}\left( 1 + z \hat \rho \right)=\prod_{n \ge0} \left(1 + z\lambda_n\right). 
\ee

Instead of using the formulation of the lattice problem in terms of the density matrix operator, we can 
introduce a quantum Hamiltonian in the standard way, 
\be
\label{rhoH}
\hat \rho=\re^{-\hat H}. 
\ee
This leads to the well-known equivalence 
between the partition function of a classical lattice gas (\ref{zellrho}) and the propagator of a quantum particle in $\ell$ units 
of discretized time (see for example \cite{parisi,kogut}). We can then write
\be
\label{exactzell}
Z_\ell =\tr\, \re^{-\ell \hat H}.
\ee
To find the Hamiltonian corresponding to the ABJM matrix model, we first write the density matrix (\ref{densitymat}) as 
\be
\label{rhopq}
\hat \rho=\re^{-{1\over 2} U(\hat q)} \re^{-T(\hat p)} \re^{-{1\over 2} U(\hat q)}.
\ee
In this equation, $\hat q, \hat p$ are canonically conjugate operators, 
\be
[\hat q, \hat p]=\ri \hbar, 
\ee
and
\be
\label{hache}
\hbar = 2 \pi k. 
\ee
This is a key aspect of this formalism: $\hbar$ is the inverse coupling constant of the gauge theory/string theory, therefore 
semiclassical or WKB expansions in the Fermi gas correspond to strong coupling expansions in gauge theory/string theory. 
The potential $U(q)$ in (\ref{rhopq}) is given by
\be
\label{upotential}
U(q)=\log \left( 2 \cosh {q\over 2} \right), 
\ee
and the kinetic term $T(p)$ is given by the same function,
\be
\label{kinabjm}
T(p)=\log\left( 2\cosh  {p \over 2} \right). 
\ee
%
%Indeed, we have 
%
%\be
%\ba
%\langle q'| \hat \rho |q\rangle&=\re^{-{1\over 2}V(q') -{1\over 2} V(q)} \int \rd p \rd p' \, \langle q'| p\rangle \langle p|\re^{-T(p)} |p'\rangle \langle p'| q\rangle\\
%&=\re^{-{1\over 2}V(q') -{1\over 2} V(q)}\int {\rd p \over 2\pi \hbar } {\re^{{\ri \over \hbar} p (q'-q)} \over 2\cosh\left( {p \over 2} \right)}\\
%&={1\over 2\pi k} \re^{-{1\over 2}V(q') -{1\over 2} V(q)} {1\over 2 \cosh \left( {q-q' \over 2 k} \right) }
%\ea
%\ee
%
%which is (\ref{densitymat}). 
The peculiar kinetic term (\ref{kinabjm}) can be regarded as a non-trivial dispersion relation interpolating between the quadratic behavior of a non-relativistic 
particle at small $p$, 
\be
T(p) \sim \log(2) + {p^2\over 8}, \qquad p \rightarrow 0, 
\ee
and the linear behavior of an ultra-relativistic particle at large $p$, 
\be
T(p) \sim {|p|\over 2} , \qquad |p| \rightarrow \infty. 
\ee

Notice that, as it is standard for Hamiltonians defined by transfer matrices at finite lattice spacing \cite{parisi,kogut}, the quantum operator $\hat H$ 
defined by (\ref{rhoH}) and (\ref{rhopq}) differs from 
\be
\label{classical}
T(\hat p) + U(\hat q)
\ee
in $\hbar$ corrections. There is a very elegant method to obtain these corrections based on the phase-space or Wigner approach to quantization. This method will be also 
extremely useful in setting the semiclassical or WKB expansion of our thermodynamic problem. 
We first recall that the Wigner transform of an operator $\hat A$ is given by (see \cite{wignerreport} for a detailed exposition of phase-space quantization)
\be
\label{wignert}
A_{\rm W}(q,p)=\int \rd q' \left\langle q-{q'\over 2}\right|\hat A \left| q+{q'\over 2}\right\rangle \re^{\ri p q'/\hbar}. 
\ee
The Wigner transform of a product is given by the $\star$-product of their Wigner transforms, 
\be
\label{starprod}
\left(\hat A \hat B\right)_{\rm W}=A_{\rm W}\star B_W 
\ee
where the star operator is given as usual by
\be
 \star=\exp\left[ {\ri \hbar \over 2} \left( {\overleftarrow{\partial}}_q {\overrightarrow{\partial}}_p  - {\overleftarrow{\partial}}_p {\overrightarrow{\partial}}_q\right) \right],
 \ee
and is invariant under linear canonical transformations. Another useful property is that
 \begin{equation}
  \Tr \hat{A}=\int\frac{\rd p \rd q }{2\pi\hbar}\,A_\wigner(q,p).
 \end{equation}
In order to calculate the $\hbar$ corrections to the Hamiltonian, 
 we consider the Wigner transform of the density matrix (\ref{rhopq}). By using (\ref{starprod}) we find, 
 \be
 \label{rhowex}
\rho_{\rm W}= \re^{-{1\over 2} U(q)}\star \re^{-T(p)} \star \re^{-{1\over 2} U(q)}.
 \ee
 Let us note that the partition function depends only on the eigenvalues $\lambda_n$ of $\hat{\rho}$ 
 (or, equivalently, on the traces $Z_\ell=\Tr\hat{\rho}^\ell$). Therefore there is the following freedom in the choice of $\hat{\rho}$:
\begin{equation}
 \hat{\rho} \rightarrow \hat{V}\hat{\rho}\hat{V}^{-1}
\end{equation} 
which translates into
\begin{equation}
\rho_\wigner(q,p) \rightarrow V_\wigner(q,p)\star \rho_\wigner(q,p)\star (V^{-1})_\wigner(q,p)
\end{equation}
after the Wigner transform. Equation (\ref{rhowex}) defines the Wigner transform of our Hamiltonian through 
 \be
 \label{rhoHW}
\rho_{\rm W}= \re_{\star}^{-H_{\rm W}}, 
\ee
where the $\star$-exponential is defined by
\be
\exp_\star(A)=1 + A + {1\over 2} A\star A + \cdots. 
\ee
The quantum Hamiltonian 
$H_\wigner$ can be computed by using the Baker--Campbell--Hausdorff formula, 
as applied to the $\star$-product. One finds,  
 \be
 \label{Hw}
 \ba
 H_{\rm W}(q,p)&=T+U+\frac{1}{12}\left[T,\left[T,U\right]_\star \right]_\star
 +\frac{1}{24}\left[U,\left[T ,U \right]_\star \right]_\star+\ldots \\
 &=T(p)+U(q)-\frac{\hbar^2}{12}\left(T'(p)\right)^2 U''(q)
 +\frac{\hbar^2}{24}\left(U'(q)\right)^2T''(p)+\CO(\hbar^4), 
\ea
\end{equation}
where we have used the fact that, at leading order in $\hbar$, the Moyal bracket is the Poisson bracket
\be
[A, B]_\star\equiv A\star B-B\star A= \ri \hbar \{ A, B\} + \CO(\hbar^2). 
\ee
Further corrections to (\ref{Hw}) can be computed to any desired order, see (\ref{Hww}) for the result at order $\CO(\hbar^4)$.

\subsection{More general Chern--Simons--matter theories}
\label{subsect_generalCSM}
\begin{figure}
\begin{center}
\includegraphics[scale=1]{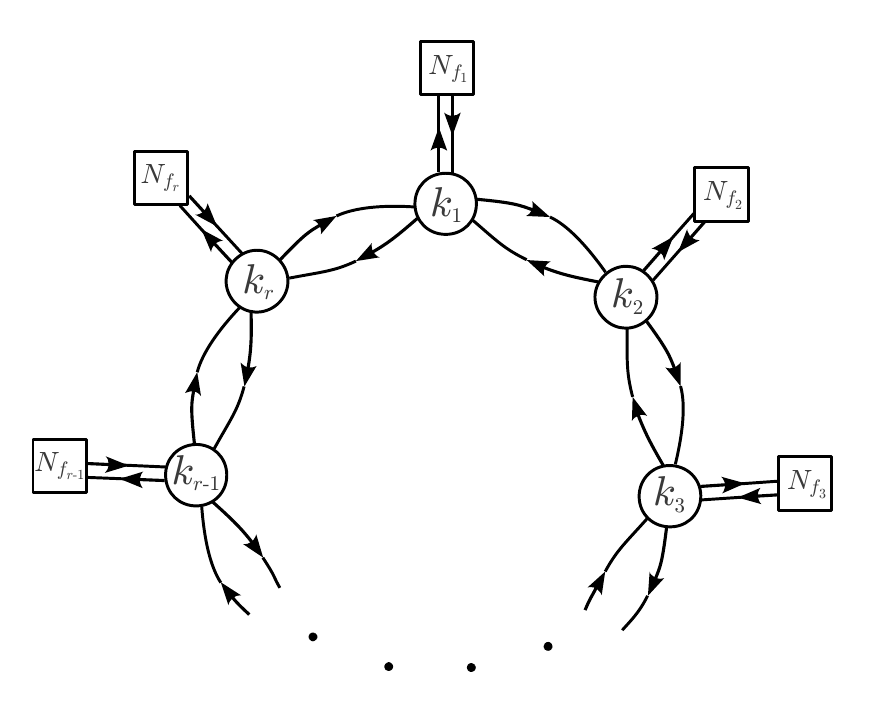}
\end{center}
\caption{A quiver with $r$ nodes forming a necklace.}
\label{fig_generalnecklace}
\end{figure}
The identification of the matrix model of ABJM theory as the partition function of a Fermi gas can be also made for more general $\CN\ge 3$ 
Chern--Simons--matter theories. 
We will set up the formalism for the necklace quivers with $r$ nodes considered in \cite{quiver1,quiver2}, and with fundamental matter in each node (see Fig.~\ref{fig_generalnecklace}). These theories are given by a 
\be
U(N)_{k_1} \times U(N)_{k_2} \times \cdots U(N)_{k_r}
\ee
Chern--Simons quiver. Each node will be labelled with the letter $a=1, \cdots, r$. There are bifundamental chiral superfields $A_{a a+1}$, $B_{a a-1}$ connecting 
adjacent nodes, and in addition we will suppose that there are $N_{f_a}$ matter superfields $(Q_a, \tilde Q_a)$ in each node, in the fundamental representation. We will write
\be
k_a=n_a k, 
\ee
and we will assume that 
\be
\label{add0}
\sum_{a=1}^r n_a=0. 
\ee

According to the general localization 
computation in \cite{kwy}, the matrix model computing the $\IS^3$ partition function of a necklace quiver is given by 
\be
Z(N)={1\over (N!)^r} \int  \prod_{a,i} {\rd \lambda_{a,i} \over 2 \pi}  {\exp \left[ {\ri n_a k\over 4 \pi}\lambda_{a,i}^2 \right] \over \left( 2 \cosh{\lambda_{a,i} \over 2}\right)^{N_{f_a}} } \prod_{a=1}^r  {\prod_{i<j} \left[ 2 \sinh \left( {\lambda_{a,i} -\lambda_{a,j} \over 2} \right)\right]^2 \over \prod_{i,j} 2 \cosh \left( {\lambda_{a,i} -\lambda_{a+1,j} \over 2} \right)}.
\ee
This matrix model is very similar to the $\hat A_{r-1}$ models considered in for example \cite{kostov,cftmm}, and one can use a very similar strategy in order to rewrite them as Fermi gases. First of all, we define a kernel corresponding to a pair of connected nodes $(a,b)$ by, 
\be
\label{kerquiv}
 K_{ab}(x',x)=\frac{1}{2\pi k}\frac{\exp\left\{\ri\frac{n_b x^2}{4\pi k}\right\}}{2\cosh\left(\frac{x'-x}{2k}\right)}\,\left[2\cosh\frac{x}{2k}\right]^{-N_{f_b}},
 \ee
 where we set $x=\lambda/k$. The grand canonical partition function corresponding to the above matrix model is defined as in (\ref{grand}). Then, 
 if we use the Cauchy identity (\ref{cauchy}), a simple generalization of the above arguments makes it possible to write it again as a Fredholm determinant (\ref{fred-det}), 
 where now \cite{kostov}
 \be
 \label{rhoquiv}
\hat \rho=\hat K_{r1}\hat K_{12}\cdots \hat K_{r-1,r}
\ee
is the product of the kernels (\ref{kerquiv}) around the quiver. Therefore, we can apply exactly the same techniques that we used before in ABJM theory. In a sense, we are 
``integrating out" $r-1$ nodes of the quiver in order to define an effective theory in the $r$-th node, but with a complicated Hamiltonian which takes into account 
the other nodes. 

This idea can be made very concrete by looking at the Wigner transform of the operator $\hat \rho$ in (\ref{rhoquiv}). We first compute the Wigner transform of the kernel (\ref{kerquiv}), 
\begin{equation}
 K^\wigner_{ab}(q,p)=\frac{1}{2\cosh\frac{p}{2}}\star\frac{\re^{\frac{\ri n_b q^2}{2\hbar}}}{\left[2\cosh\frac{q}{2k}\right]^{N_{f_b}}}
\end{equation} 
where the $\hbar$ in the $\star$ product is given again by (\ref{hache}). Let us note that
\begin{equation}
 \re^{\frac{\ri nq^2}{2\hbar}}\star f(p)\star \re^{-\frac{\ri nq^2}{2\hbar}}=
f\left(\re^{\frac{\ri nq^2}{2\hbar}}\star p\star \re^{-\frac{\ri n q^2}{2\hbar}}\right)=
f\left(\re^{\ad_\star\left[\frac{\ri nq^2}{2\hbar}\right]} p\right)=f(p-nq),
\end{equation} 
where we used that
\begin{equation}
 \left[q^2,p\right]_\star=2\ri\hbar q.
\end{equation}
We obtain then, for the Wigner transform of the density operator (\ref{rhoquiv})
\be
\ba
 \rho_\wigner(q,p)&=\frac{1}{2\cosh\frac{p}{2}}\star\frac{1}{\left[2\cosh\frac{q}{2k}\right]^{N_{f_1}}}\star
\frac{1}{2\cosh\frac{p-n_1q}{2}}\star \\
& \frac{1}{\left[2\cosh\frac{q}{2k}\right]^{N_{f_2}}}\star
\frac{1}{2\cosh\frac{p-(n_1+n_2)q}{2}}\star\frac{1}{\left[2\cosh\frac{q}{2k}\right]^{N_{f_3}}}\star\\
& \cdots \star
\frac{1}{2\cosh\frac{p-(n_1+\cdots+n_{r-1})q}{2}}\star\frac{1}{\left[2\cosh\frac{q}{2k}\right]^{N_{f_r}}}
\ea
\label{rho_wigner_quiv}
\ee
where we used (\ref{add0}). For necklace theories without fundamental matter this is simply
\begin{equation}
 \rho_\wigner(q,p)=\frac{1}{2\cosh\frac{p}{2}}\star
\frac{1}{2\cosh\frac{p-n_1q}{2}}\star
\frac{1}{2\cosh\frac{p-(n_1+n_2)q}{2}}\star
\cdots \star
\frac{1}{2\cosh\frac{p-(n_1+\cdots+n_{r-1})q}{2}}.
\end{equation} 
In particular, for the ABJM necklace $(-k,k)$ with fundamental matter $N_{f_1}=N_{f_2}=N_f$ first considered in \cite{hk,gj,taka}, we have
\begin{equation}
 \rho_\wigner(q,p)=\frac{1}{2\cosh\frac{p}{2}}\star\frac{1}{\left[2\cosh\frac{q}{2k}\right]^{N_f}}\star
\frac{1}{2\cosh\frac{p+q}{2}}\star\frac{1}{\left[2\cosh\frac{q}{2k}\right]^{N_f}}.
\label{rho_wigner_matter}
\end{equation} 
If we perform a canonical transformation 
\be
p\rightarrow-q, \qquad q\rightarrow p+q
\ee
and we conjugate by $\left[2\cosh\frac{q}{2}\right]^{1/2}$ to obtain a symmetric kernel, we get the equivalent representation, 
\begin{equation}
\rho_\wigner(q,p)=\frac{1}{\left[2\cosh\frac{q}{2}\right]^{1/2}}\star
\frac{1}{\left[2\cosh\frac{p+q}{2k}\right]^{N_f}}\star
\frac{1}{2\cosh\frac{p}{2}}\star\frac{1}{\left[2\cosh\frac{p+q}{2k}\right]^{N_f}}\star
\frac{1}{\left[2\cosh\frac{q}{2}\right]^{1/2}}
\end{equation} 
which, for $N_f=0$, agrees with the result (\ref{rhowex}). In this way, we have reduced the general necklace quiver theory to an ideal Fermi gas whose one-particle 
quantum Hamiltonian is defined by the above density matrices through (\ref{rhoHW}). 

Notice that, in general, the density operators $\hat \rho$ are not Hermitian, and correspondingly 
$H_{\rm W}$ is generally not real. This reflects the fact that the free energy on the three-sphere of these CSM theories is in general complex. 

\sectiono{Thermodynamic limit}

It is well-known that the thermodynamic limit of an ideal quantum gas can be evaluated by treating the one-particle problem in the semiclassical or WKB approximation. Moreover, the $1/N$ corrections to the thermodynamic limit can be obtained by 
studying the quantum corrections to the semiclassical limit. In this section we will present general results about the thermodynamic limit and we will illustrate them in ABJM theory. More general theories will be considered in section \ref{genCSM}. 

\subsection{The thermodynamic limit of ideal Fermi gases}

In the following we will need several standard results in the analysis of ideal quantum gases. The 
distribution operator at zero temperature is given by, 
\be
\label{Nop}
\hat n(E)=\theta(E -\hat H)
\ee
where $\theta(x)$ is the Heaviside step function. The trace of this operator gives the function $n(E)$, counting the number of 
eigenstates whose energy is less than $E$:
\be
\label{Noper}
n(E)=\tr \, \hat n(E) =\sum_n  \theta(E -E_n).
\ee
Notice that 
\be
E_n =-\log \lambda_n, 
\ee
where $\lambda_n$ are the eigenvalues (\ref{eigenv}) of the density matrix. 
The density of eigenstates is defined by 
\be
\rho(E) ={\rd n(E) \over \rd E}=\sum_n \delta(E-E_n). 
\ee
The one-particle canonical partition function is then given by the standard formula, 
\be
Z_\ell =\int_0^{\infty} \rd E \, \rho(E) \, \re^{-\ell E}, 
\ee
while the grand-canonical potential of the $N$ particle system is given by 
\be
\label{Jrho}
J(\mu)=\int_0^{\infty} \rd E \, \rho(E) \, \log\left(1+ z\re^{-E} \right). 
\ee

Let us now consider the thermodynamic limit of the system, when $N \rightarrow \infty$. In this regime, the behavior of the system is semiclassical 
and the spectrum of the 
one-particle Hamiltonian is encoded in the functions $n(E)$, $\rho(E)$. The thermodynamic limit is governed by the 
behavior of these functions as $E \gg 1$. We notice that, if 
\be
\label{Nlimit}
n(E) \approx C E^s, \qquad E \gg 1, 
\ee
then the grand-canonical potential is given by
\be
 J(\mu) \approx
 sC\int\limits_0^\infty \log \left(1+z\re^{-E}\right)E^{s-1}\rd E=-C\Gamma(s+1)\Li_{s+1}(-\re^{\mu}),
\end{equation} 
where ${\rm Li}_s$ is the usual polylogarithm function. The number of particles is related to the chemical potential by
\begin{equation}
 N(\mu) \approx C\Gamma(s+1)\Li_{s}(-\re^{\mu}), 
 \ee
 and large $N$ corresponds to large $\mu$. In this regime, we have
\begin{equation}
\label{Nmu}
 J(\mu)\approx \frac{C}{s+1}\mu^{s+1}, \qquad N(\mu)\approx C\mu^s\,.
\end{equation} 
The second equation defines $\mu$ as function of $N$, and we deduce from (\ref{freethermo}) that the canonical free energy is given, as $N \rightarrow \infty$, by
\begin{equation}
\label{freeenergy}
 F(N)\approx -\frac{s}{s+1} C^{-1/s} N^\frac{s+1}{s}\,.
\end{equation} 
These formulae should be familiar from the elementary theory of ideal quantum gases. For example, the textbook 
ideal Fermi gas in three dimensions has $s=3/2$. 

To determine the value of $s$ for a given system we notice that, in the semiclassical limit, the trace is replaced 
by an integral over phase space
\be
\tr \rightarrow \int {\rd q \rd p \over 2 \pi \hbar} 
\ee
which gives the standard semiclassical formula 
\be 
\label{semivol}
n(E) \approx \int {\rd p \rd q \over 2 \pi k} \theta (E -H(q,p))={\Vol (E) \over 2\pi \hbar},
\ee
i.e. the number of eigenstates is just given by the volume of phase space. The surface
\be
\label{fermi}
H (q,p)=E
\ee
is just the Fermi surface of the system. For a one-dimensional ideal gas whose one-particle Hamiltonian is of the form 
\be
\label{genH}
H \sim A |p|^{\alpha} + B |q|^{\beta}
\ee
we have 
\be
\label{generals}
s={1\over \alpha} +{1\over \beta}. 
\ee
This will be useful later on.

\subsection{A simple derivation of the $N^{3/2}$ behavior in ABJM theory}

\FIGURE{
\includegraphics[height=6cm]{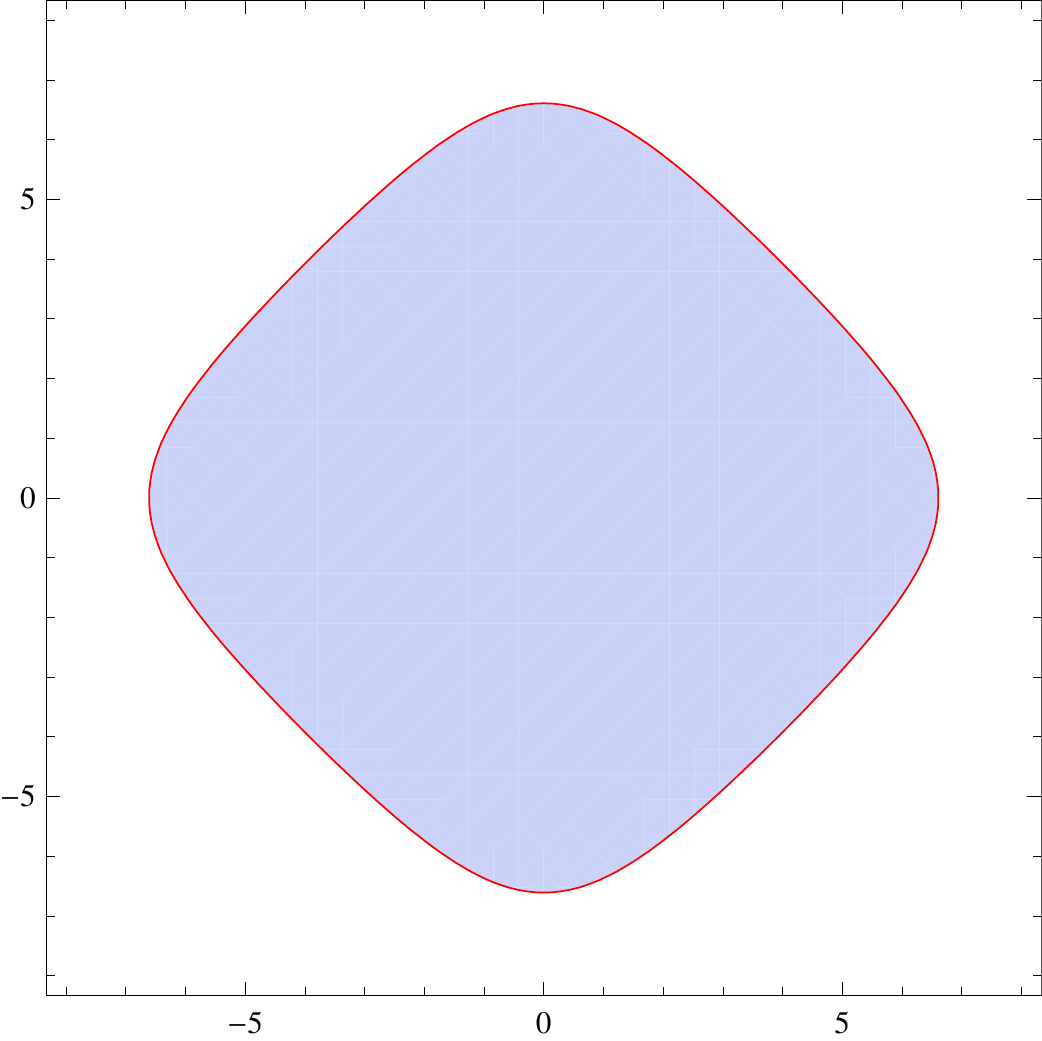} \qquad 
\includegraphics[height=6cm]{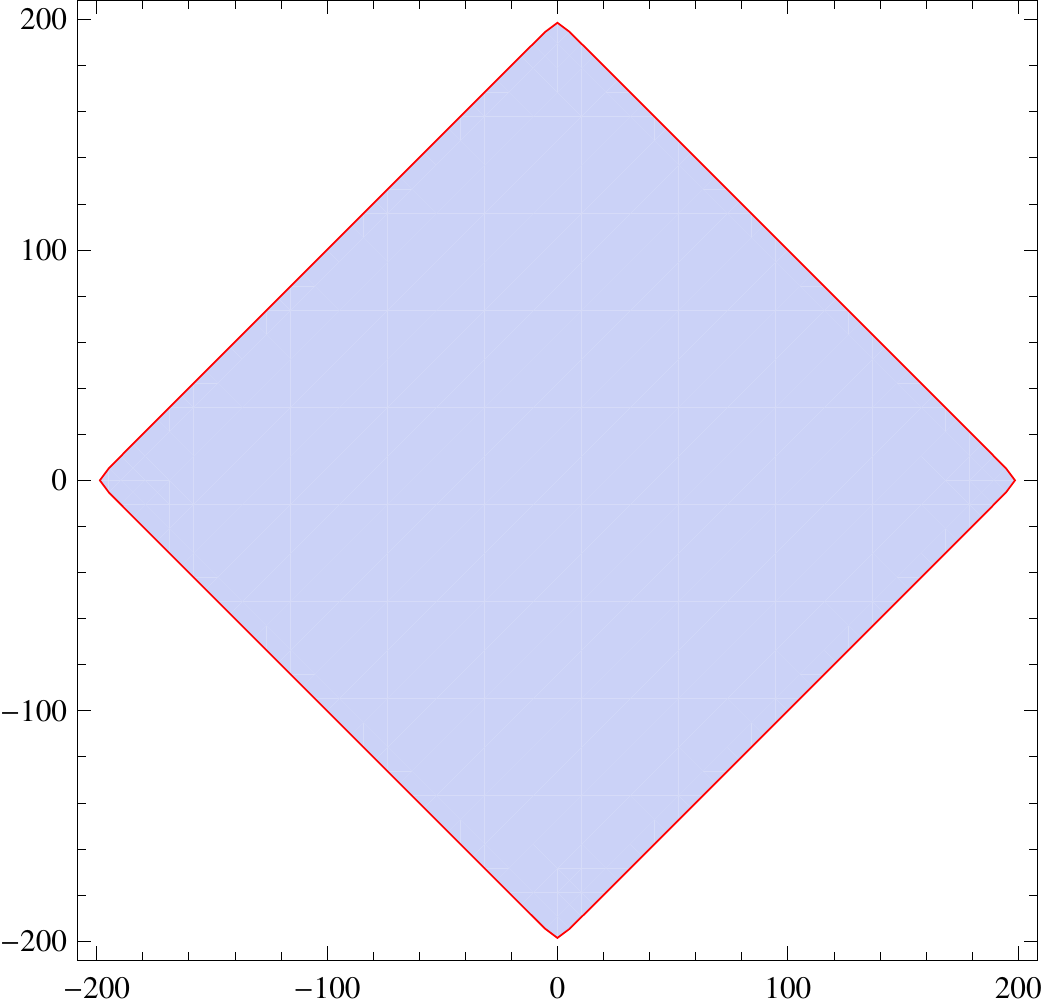}
\caption{The Fermi surface (\ref{clfermi}) for ABJM theory in the $q$-$p$ plane, for $E=4$ (left) and $E=100$ (right). When the energy is large, the Fermi surface approaches the polygon (\ref{polygon}).}
\label{fermisurface}
}
We can now study the thermodynamic limit of the partition function of ABJM theory. In this case, the Hamiltonian appearing in the semiclassical formula 
(\ref{semivol}) is just given by the classical counterpart of (\ref{classical}),  
\be
\label{trulyclas}
H_{\rm cl}(q,p)=T(p) + U(q)=\log\left(2 \cosh {p \over 2} \right) +\log\left(2 \cosh {q \over 2} \right).
\ee
Here we have neglected the $\hbar$ corrections appearing in $H_{\rm W}$. It is easy to show that the minimum energy is 
\be
\label{minen}
E_0=2 \log 2, 
\ee
which corresponds to the maximal eigenvalue of the density matrix
\be
\label{leadlam}
\lambda_0={1\over 4}. 
\ee
This is the semiclassical value given by the leading WKB approximation, and it will be corrected quantum-mechanically.
In the large $E$ regime, the discrete spectrum ``condenses" along a cut in the complex plane, and $\lambda_0$ signals the endpoint of the cut.

In order to proceed with the analysis of the thermodynamic limit, we should determine the Fermi surface
\be
\label{clfermi}
H_{\rm cl}(q,p)=E
\ee
controling the density of 
eigenvalues. We show the shape of this surface in \figref{fermisurface} for $E=4$ (left) and $E=100$ (right). 
It is clear that in the thermodynamic limit, when $E$ is large, the surface can be approximated by considering the 
values of $U(q)$, $T(p)$ for $q,p$ large. In this regime we have
\be
U(q) \approx {|q| \over 2}, \quad |q| \rightarrow \infty, \qquad T(p) \approx {|p|\over 2}, \quad |p| \rightarrow \infty, 
\ee
so that (\ref{clfermi}) is approximately given by 
\be
\label{polygon}
|q|+|p|=2 E,
\ee
as it is manifest in the graphic on the right in \figref{fermisurface}. From (\ref{genH}) and (\ref{generals}) we deduce that 
\be
s=2. 
\ee
Since
\be
{\rm Vol}(E) \approx 8 E^2,
\ee
the number of states is given by 
\be
\label{semisimple}
n(E)\approx {2  \over \pi^2 k} E^2. 
\ee
By comparing with (\ref{Nlimit}), we find 
\be
C={2\over \pi^2 k}.
\ee
The equation (\ref{freeenergy}) gives immediately
\be
\label{n32}
 F(N)\approx -\frac{\pi\sqrt{2k}}{3}N^{3/2}\,.
\end{equation} 
This is exactly the result first found in \cite{dmp} using the 't Hooft expansion of the matrix model. The derivation presented here 
is however completely elementary, and relies on basic notions of quantum Statistical Mechanics: the $3/2$ scaling of the number of degrees of freedom is nothing 
but the scaling of the free energy of an ultrarelativistic gas of one-dimensional fermions in a linearly confining potential. 
No matrix model techniques are needed. In this sense, our 
derivation is even simpler than the one presented in \cite{hklebanov}, which required some detailed analysis of the eigenvalue interaction in the matrix integral. 

We would like to emphasize that the above result (\ref{n32}) provides the right large $N$ behavior of the system at finite $k$. 
This is because the true expansion parameter in the semiclassical expansion is $\hbar/E$, which is small for large $E$ even at finite $\hbar$. This can be proved rigorously for 
some spectral problems defined by kernels of the form (\ref{densitymat}) \cite{widom}, and we will verify it in section \ref{qcorr} by a detailed analysis of the WKB expansion.

\subsection{Large $N$ corrections} 
\label{largencorabjm}
One advantage of the statistical-mechanical framework presented here is that it makes it possible to compute corrections to the thermodynamic limit in a systematic way. 
To start the study of these corrections, we now look at the thermodynamics of the Fermi gas of ABJM theory in the semiclassical approximation, but taking into account the 
exact value of the volume of phase space (i.e. we go beyond the polygonal approximation in (\ref{polygon})). As expected, this gives sub-leading and exponentially suppressed 
corrections at large $N$. 

The computation of the exact volume is equivalent to computing 
all the $Z_\ell$ exactly in the semiclassical approximation, and resumming the resulting series (\ref{jsum}). Using that
\be
\label{coshint}
\int_{-\infty}^\infty\frac{\rd\xi}{\left(2\cosh{\xi \over 2} \right)^\ell}=\frac{\Gamma^2(\ell/2)}{\Gamma(\ell)}
\ee
we find
 \begin{equation}
 \label{firstint}
 Z_\ell\approx  {1\over \hbar} Z_\ell^{(0)}, 
 \ee
 where 
 \be
Z_\ell^{(0)}=  \int\frac{\rd p \rd q}{2\pi}\re^{-\ell H_{\rm cl}(q,p)}
 =\frac{1}{2\pi}\frac{\Gamma^4(\ell/2)}{\Gamma^2(\ell)}\,.
\end{equation}
Therefore, 
\begin{equation}
 J(\mu)\approx \frac{1}{k}J_0(\mu)
\end{equation} 
where
\begin{equation}
\label{exactJ0}
\ba
 J_0(\mu)&=-\sum_{\ell=1}^\infty\frac{(-z)^{\ell}}{4\pi^2}\frac{\Gamma^4(\ell/2)}{\ell \Gamma^2(\ell)}\\
 &=
 \frac{1}{4} z \, _3F_2\left(\frac{1}{2},\frac{1}{2},\frac{1}{2};1,\frac{3}{2};\frac{z^2}{16}\right)-\frac{z^2}{8 \pi ^2} \, _4F_3\left(1,1,1,1;\frac{3}{2},\frac{3}{2},2;\frac{z^2}{16}\right).
 \ea
\end{equation}
This function has a branch cut in the $z$-plane at $(-\infty, -4]$. This is expected: indeed, from (\ref{fred-det}) there should be a cut starting at 
\be
z=-\lambda^{-1}_0=-4,
\ee
indicating the condensation of eigenvalues for the one-particle density matrix. The function (\ref{exactJ0}) has the following asymptotics for large $\mu$, 
\begin{equation}
 J_0(\mu)=\frac{2 \mu ^3}{3 \pi ^2}+\frac{\mu }{3}+\frac{2 \zeta (3)}{\pi ^2}+J_0^{\rm np}(\mu). 
\end{equation}
The leading, cubic term in $\mu$, is the responsible for the behavior (\ref{n32}). The subleading term in $\mu$ gives a correction of order $N^{1/2}$ to the leading 
behavior (\ref{n32}). The last, non-perturbative term involves an infinite power series of exponentially small corrections in $\mu$. They have the structure, 
\be
J^{\rm np}_0(\mu)=\sum_{\ell=1}^{\infty} \left(a_{0,\ell} \mu^2 + b_{0,\ell} \mu + c_{0,\ell} \right) \re^{-2 \ell \mu}.
\ee
Explicitly, one finds for the very first orders
\be
\ba
J^{\rm np}_0(\mu)&={2 \over 3 \pi^2} \left(6-\pi^2 + 6 \mu - 6 \mu^2\right) \re^{-2 \mu}+ {1 \over 2 \pi^2} \left(25-6 \pi^2 -66 \mu - 36 \mu^2\right) \re^{-4 \mu}\\
&+\CO\left(\mu^2 \re^{-6 \mu}\right).
\ea
\ee
The non-perturbative part of the grand potential leads to exponentially small corrections in $N$ in the canonical free energy. In fact, using (\ref{Nmu}) we find 
that, once evaluated at the saddle-point, 
\be
\exp\left(-2\mu\right) \approx \exp\left(-{\sqrt{2}} \pi k^{1/2} N^{1/2} \right). 
\ee
This is precisely the action for membrane instantons (\ref{membranea}) found in \cite{dmpnp} as large $N$ instantons of the matrix model in the 't Hooft expansion. We conclude that the exponentially small 
corrections in $\mu$, which in this approach appear already in the semi-classical approximation, correspond in fact to non-perturbative corrections in the genus expansion, and should be identified as membrane instanton contributions. 

As mentioned before, the calculation of these exponentially small corrections to the grand-canonical potential is equivalent to the exact calculation of the volume (\ref{semivol}) of classical phase space. To see this, we notice that we can write this volume as a period of the one-form $p \rd q$ along the curve (\ref{fermi})
\be
\label{volE}
 \Vol(E)=\oint\limits_{H_{\rm cl}(q,p)=E} p \rd q. 
\ee
 This period vanishes at the point $E=E_0$. It turns out that its exact value is given by a Meijer function,
 \begin{equation}
 \label{volexact}
 \Vol(E)=\frac{\re^E}{\pi}  G_{3,3}^{2,3}\left(\frac{\re^{2E}}{16}\left|
\begin{array}{c}
 \frac{1}{2},\frac{1}{2},\frac{1}{2} \\
 0,0,-\frac{1}{2}
\end{array}
\right.\right)-4\pi^2.
\end{equation}
This leads to the following large $E$ expansion of the number of states, 
\begin{equation}
\label{asymstates}
n(E) =\frac{2E^2}{\pi^2 k}-\frac{1}{3k}+\CO(E\re^{-2E})+\mathcal{O}(\hbar),
\end{equation} 
where the first term agrees of course with the semiclassical calculation at large $E$ done before. One can then check that the expression (\ref{Jrho}) for the grand-canonical potential 
reproduces (\ref{exactJ0}), once the density obtained from (\ref{volexact}) is used. 

\subsection{Relation to previous results}
The semiclassical limit of the one-particle Hamiltonian turns out to be closely related to the planar limit of ABJM theory 
studied in \cite{mp,dmp,dmpnp}. 

First of all, the semiclassical quantization of the one-dimensional problem leads to the Fermi surface 
\be
\label{fermiE}
T(p)+ U(q) =E
\ee
which is in fact a {\it curve} in phase space. Let us now make the following change of variables, 
\be
x={q \over 2}+{p \over 2}, \qquad y=p + \pi \ri, 
\ee
which, up to an overall constant, preserves the form $p \rd q$. 
In terms of the exponentiated variables
\be
X=\re^{q/2+p/2}, \qquad Y=-\re^p. 
\ee
the Fermi surface (\ref{fermiE}) reads
\be
\label{sc}
 Y+\frac{X^2}{Y}-X^2+\ri\kappa X-1=0, 
 \ee
 where 
 \be
 \ri \kappa=\re^E. 
 \label{kappa_energy}
 \ee
 The curve (\ref{sc}) is nothing but the spectral curve of the ABJM matrix model written down in for example \cite{dmp}. 
 The minimal energy $E_0$ given in (\ref{minen}) corresponds to the 
 conifold point $\kappa=4\ri$ studied in detail in \cite{dmp}. The volume of phase space, which as we remarked after (\ref{volE}) is a vanishing period at $E=E_0$, 
 is actually proportional to the conifold period studied in \cite{dmpnp}. Finally, the large energy limit, in which the Fermi surface becomes a polygon, is nothing but 
 the tropical limit of the spectral curve, studied in \cite{cmp}. 
 
\sectiono{Quantum corrections}
\label{qcorr}
In the previous section we have recalled the semiclassical limit of ideal Fermi gases, and we have studied in detail the case of ABJM theory. 
We now study the corrections to the semiclassical limit in a systematic and general way. These corrections lead to a power series in $\hbar^2\propto k^2$ for 
the grand-canonical partition function. As we will see, only the first $\hbar^2$ correction contributes to the asymptotic series in $1/N$ of the canonical 
free energy, up to an additive function of $k$ but independent of $N$. This means that we can compute the {\it full} series of $1/N$ corrections to the original matrix model partition function, up to an overall, $N$-independent constant. However, the exponentially small terms in $\mu$ appearing in $J(\mu)$ receive corrections to all orders in $\hbar^2$. 

\subsection{Quantum-corrected Hamiltonian and Wigner--Kirkwood expansion}
\label{firstqc}

There are two sources of $\hbar$ corrections in the one-body problem appearing in our Fermi gas formulation. The first one appears already in the Hamiltonian 
$\hat H$: when we compute $\hat H$ starting from (\ref{rhopq}), the non-commutativity of the operators in (\ref{rhopq}) leads to $\CO(\hbar)$ corrections to (\ref{classical}). This first source 
of corrections is nicely encoded in the Wigner transform (\ref{Hw}). Another source 
of corrections is due to the standard semiclassical expansion of the density of eigenvalues. We now present a formalism to treat in a systematic way both 
types of corrections. This formalism is a generalization of the standard Wigner--Kirkwood $\hbar$ expansion \cite{wigner,kirkwood} in quantum statistical mechanics, and it incorporates general, $\hbar$-dependent Hamiltonians. As in section \ref{rewriting}, the formalism 
is most conveniently formulated in the phase-space approach to quantization, and it has been 
developed in the context of many-body physics. The most elegant presentation is due to Voros \cite{voros,gramvoros} (see also \cite{centelles}).

 Let $\hat H$ be the Hamiltonian of a one-particle, one-dimensional 
 quantum system, and let $H_{\rm W}$ be its Wigner transform. We would like to compute systematically the 
 $\hbar$ expansion of the canonical partition function and of the density of states. Following \cite{gramvoros} we notice that it is possible 
 to expand any function $f(\hat H)$ of $\hat H$ around $H_{\rm W}(q,p)$, which is a $c$-number. This gives, 
\be
f(\hat H) = \sum_{r\ge 0} {1\over r!} f^{(r)}( H_{\rm W}) \left( \hat H -H_{\rm W}(q,p)\right)^r. 
\ee
The semiclassical expansion of this object is obtained simply by evaluating its Wigner transform, and we obtain
\be
f(\hat H)_{\rm W} = \sum_{r\ge 0} {1\over r!} f^{(r)}\left( H_{\rm W} \right)\CG_r  
\ee
where
\be
\label{Gr}
\CG_r=\left[ \left( \hat H -H_{\rm W}(q,p)\right)^r \right]_{\rm W}
\ee
and the Wigner transform is evaluated at the same point $q,p$. Of course, one has 
\be
\CG_0=1, \qquad \CG_1=0,
\ee
and the quantities $\CG_r$ for $r\ge 2$ can be computed again by using (\ref{starprod}). They have an $\hbar$ expansion of the form 
\be
\label{GrEx}
\CG_r=\sum_{n\ge \left[\frac{r+2}{3}\right]} \hbar^{2n} \CG_r^{(n)},  \qquad r\ge 2.
\ee
This means, in particular, that to any order in $\hbar^2$, only a finite number of $\CG_r$'s are involved. 
One finds, for the very first orders \cite{gramvoros, centelles},
\be
\ba
\CG_2&=-{\hbar^2 \over 4} \left[ {\partial^2 H_{\rm W} \over \partial q^2}{\partial^2 H_{\rm W} \over \partial p^2}-\left( {\partial^2 H_{\rm W} \over \partial q \partial p}\right)^2 \right] +\CO(\hbar^4), \\
 \CG_3&=-{\hbar^2 \over 4} \left[\left( {\partial H_{\rm W} \over \partial q} \right)^2  {\partial^2 H_{\rm W} \over \partial p^2}+ 
 \left( {\partial H_{\rm W} \over \partial p} \right)^2  {\partial^2 H_{\rm W} \over \partial q^2}-2  {\partial H_{\rm W} \over \partial q} {\partial H_{\rm W} \over \partial p}{\partial^2 H_{\rm W} \over \partial q \partial p}  \right]+\CO(\hbar^4). 
 \ea
 \ee
One can then apply this method to compute the semiclassical expansion of any function of the Hamiltonian operator. 
For example, when applied to (\ref{Noper}), one finds, 
\be
\hat n(E)_{\rm W}= \theta(E-H_{\rm W})+ \sum_{r=2}^{\infty} {1\over r!} \CG_r \delta^{(r-1)}(E-H_{\rm W}), 
\ee
therefore
\be
\label{exactdens}
n(E)=\int_{H_{\rm W}(q,p) \le E} {\rd q \rd p \over 2 \pi \hbar} +  \sum_{r=2}^{\infty} {1\over r!} \int {\rd q \rd p \over 2 \pi \hbar} \CG_r \delta^{(r-1)}(E-H_{\rm W}). 
\ee
When applied to the canonical density matrix at inverse temperature $\beta$, one finds, 
\be
\label{genwk}
\left( \re^{-\beta \hat H} \right)_{\rm W}=\left( \sum_{r=0}^{\infty}{ (-\beta)^r  \over r!} \CG_r \right) \re^{-\beta H_{\rm W}}. 
\ee
The standard Wigner--Kirkwood $\hbar$ expansion of the canonical partition function \cite{wigner,kirkwood} is just a particular case of (\ref{genwk}) when the Hamiltonian is
 \be
 \label{standardH}
 \hat H ={\hat p^2 \over 2} + U(\hat q). 
 \ee

Let us now apply this formalism to our case. First of all, the quantum-corrected Hamiltonian is given by a power series in $\hbar$ of the form, 
 \be
 H_{\rm W} =\sum_{n\ge 0} \hbar^{2n}H^{(n)}_{\rm W}.
 \ee
At leading order we find of course the classical Hamiltonian (\ref{trulyclas}), 
\be
H^{(0)}_{\rm W}=T(p) + U(q), 
\ee
the $\CO(\hbar^2)$ term is written down in (\ref{Hw}), and the $\CO(\hbar^4)$ term can be found in (\ref{Hww}). 
The one-particle canonical partition function can then be computed as a power 
series in $\hbar$, 
\be
Z_\ell ={1\over \hbar} \sum_{n=0}^{\infty} Z_\ell^{(n)} \hbar^{2n}, 
\ee
where 
\be
Z_\ell^{(0)}= \int {\rd q \rd p \over 2 \pi} \re^{-\ell H_{\rm cl}}
\ee
is the classical limit. The expansion is obtained by grouping $\hbar^2$ corrections in the expression 
\be
Z_\ell ={1\over \hbar}\sum_{r\ge 0} {(-\ell)^r \over r!} \int {\rd q \, \rd p \over 2 \pi} \CG_r \re^{-\ell H_{\rm W}}. 
\ee
The power series in $\hbar$ for $Z_\ell$ leads to the following power series in $k$ for $J(\mu)$, 
\be
J(\mu)={1\over k} \sum_{n=0}^{\infty} J_n(\mu) k^{2n},
\ee
where
\be
J_n(\mu) =-(2 \pi)^{2n-1} \sum_{\ell=1}^{\infty} {(-z)^\ell \over \ell} Z_\ell^{(n)}. 
\ee

As an illustration of the above, general considerations, we will now calculate the first, $\hbar^2$ correction to the semiclassical 
result of ABJM theory obtained in section \ref{largencorabjm}. Using the formulae above, we find
\be
\ba
Z_\ell^{(1)}&= \int  {\rd q \rd p \over 2 \pi } \re^{-\ell H_{\rm cl}} \left\{ -\ell H_{\rm W}^{(1)} + {\ell^2 \over 2} \CG_2^{(1)} 
-{\ell^3 \over 6}  \CG_3^{(1)} \right\}\\
&=  -\ell  \int  {\rd q \rd p \over 2 \pi } \re^{-\ell H_{\rm cl}} \left[ {1\over 24} (U'(q))^2 T''(p) -
{1\over 12} (T'(p))^2 U''(q)\right] \\
&  +  \int {\rd q \rd p \over 2 \pi } \re^{-\ell H_{\rm cl}} \left\{ {\ell^3 \over 24} \left[ (U'(q))^2 T''(p) +U''(q) (T'(p))^2 \right] -{ \ell^2 \over 8} U''(q)T''(p) \right\}.
\ea
\ee
To evaluate these coefficients, we need the integral appearing in (\ref{coshint}), as well as 
\be
\int_{-\infty}^\infty\rd\xi \frac{\tanh^2 (\xi/2)}{\left(2\cosh(\xi/2)\right)^\ell}={\Gamma^2(\ell/2) \over  \Gamma(\ell)} - 4 {\Gamma^2(\ell/2+1) \over  \Gamma(\ell+2)}.
\ee
We then find, 
\be
\ba
\label{zellcor}
Z_\ell^{(1)}&= {\ell \over 48 \pi}(2\ell^2+1) \left[ {\Gamma^2(\ell/2+1)\Gamma^2(\ell/2)  \over 4 \Gamma(\ell+2) \Gamma(\ell)} -  {\Gamma^4(\ell/2+1) \over  \Gamma^2(\ell+2)}\right]-{\ell^2 \over 16 \pi} {\Gamma^4(\ell/2+1) \over  \Gamma^2(\ell+2)}.
\ea
\ee
From (\ref{zellcor}) one can compute $J_1(z)$ in closed form. Let us introduce the function 
\be
\ba
f(z)&= {~}_3F_2\left(1,1,1;\frac{3}{2},\frac{3}{2};\frac{z^2}{16}\right)-{z^2 \over 24} \,
   _3F_2\left(1,1,2;\frac{3}{2},\frac{5}{2};\frac{z^2}{16}\right)\\
   &+{1\over z}
   \left(-2 \pi  E\left(\frac{z}{4}\right)-z+\pi ^2\right)
   \ea
   \ee
 where $E(k)$ is the complete elliptic integral of the second kind with modulus $k$. 
Then, one finds
\be
J_1(\mu)= {1 \over 24} \left\{ f(z) -\left(z {\partial \over \partial z} \right)^2 f(z)\right\}. 
\ee
The asymptotic expansion of the above function at large $\mu$ is given by
\be
\label{fz}
 f(z) -\left(z {\partial \over \partial z} \right)^2 f(z)= \mu -2 + \CO\left(\mu^2\re^{-2 \mu}\right).
 \ee
 Therefore, we find, at next-to-leading order in $k$, the following expression for the grand canonical potential of ABJM theory, 
 \be
 \label{JABJM}
 J_{\rm ABJM}(\mu) \approx {2 \mu^3 \over 3 k \pi^2} + \mu  \left( {1\over 3k} +{k \over 24}\right) + {2 \zeta(3) \over \pi^2 k } -{k\over 12} +\CO\left(\mu^2\re^{-2 \mu}\right).
 \ee
Notice that the non-perturbative corrections in $\mu$ to (\ref{fz}) involve only {\it even} powers of $z$. This is consistent with their interpretation as membrane instantons. 

\subsection{General structure of quantum corrections} 
\label{subsect_generalcorr}
As we mentioned above, we can compute the quantum corrections to $J(\mu)$ either by working out the corrections to the $Z_\ell$ integrals, or by working out the corrections to the function $n(E)$. In order to understand the general structure of these corrections for a Fermi gas, the second point of view is more convenient. In this 
section we will analyze this general structure in detail, and we will make a precise connection between the structure of $n(E)$ and the expected Airy function 
behavior. 

First of all, we have to understand more precisely the relationship between the structure of $n(E)$ and the structure of $J(\mu)$. Let us write the density function $n(E)$ 
in the form, 
\be
\label{nEst}
n(E)= C E^2 + n_0 + n_{\rm np}(E), 
\ee
where the last term has the following asymptotics at infinity,
\be
\label{nnpas}
n_{\rm np}(E)= \CO(E\re^{-E}), \quad E\rightarrow\infty.
\ee
We know from (\ref{asymstates}) that this is indeed the case at leading order in $k$ for ABJM theory and in the next subsection we will show that quantum corrections do not spoil this behavior. Notice that, since all eigenvalues of our Hamiltonian are positive, we must have 
\be
n(0)=0, 
\ee
therefore
\be
\label{zerocond}
n_{\rm np}(0)=-n_0. 
\ee
If we now plug (\ref{nEst}) in (\ref{Jrho}) we find, 
\be
\ba
J(\mu)&=\int\limits_0^\infty \rd E \, n'(E)\log(1+z\re^{-E})\\ &=
-2C \Li_3(-z)+\mu \int\limits_0^\infty \rd E \,n_{\rm np}'(E)-\int\limits_0^\infty
\rd E \,n_{\rm np}'(E) E +\int\limits_0^\infty \rd E \,n_{\rm np}'(E) \log(1+ \re^E/z).
\ea
\ee
The second integral gives
\begin{equation}
\int_0^\infty \rd E\, n_{\rm np}'(E)=-n_{\rm np}(0)=n_0,
\end{equation}
where we used (\ref{zerocond}). The last term can be calculated as
\begin{equation}
 \int\limits_0^\infty \rd E\, n_{\rm np}'(E)\log(1+ \re^E/z)=n_0\log(1+ 1/z)-\int\limits_0^\infty \rd E \frac{n_{\rm np}(E)}{1+z \re^{-E}},                                        
\end{equation} 
and both terms are non-perturbative in $\mu$. Indeed, 
\begin{equation}
\int\limits_0^\infty \rd E \frac{n_{\rm np}(E)}{1+z \re^{-E}} \sim \int\limits_0^\infty \rd E \frac{E \re^{- E}}{1+z \re^{-E}}=\CO\left(\mu \, \re^{-\mu}\right).
\end{equation}
Then, by using the standard asymptotics of the trilogarithm 
\be
{\rm Li}_3(-z) = -{\mu^3\over 6} -{\pi^2 \over 6} \mu +\CO\left(\re^{-\mu}\right), 
\ee
we deduce the following asymptotic expansion of $J(\mu)$ for large $\mu$:
\begin{equation}
\label{Jden}
 J(\mu)=\frac{C}{3}\mu^3+B\mu+A+J_\np(\mu)\,,
\end{equation} 
where
\begin{equation}
\label{shift_rel}
\ba
 B&=n_0+{\pi^2 C \over 3},\\
 A&=-\Tr' \hat H\equiv -\int\limits_0^\infty \rd E \, E\,  n_{\rm np}'(E),
 \ea
\end{equation}
and 
\be
J_\np(\mu)=\CO\left(\mu \,  \re^{-\mu}\right). 
\ee
Notice that $A$ is a non-trivial function of $k$, but it doesn't depend on $\mu$. If we now plug this in (\ref{muint}), we find immediately
\begin{equation}
\label{airydens}
 Z(N)=C^{-1/3}\re^{A}\Ai\left[C^{-1/3}(N-B)\right]+Z_\np(N),
\end{equation} 
where the last term is non-perturbative in $N$. 

We then see that, if we are able to derive the structural results (\ref{nEst}) and (\ref{nnpas}) for the density of states of a given theory, 
the conjecture \ref{airycon} for the M-theory expansion is proved. 
In fact, so far we have not specified in which regime we are working in $k$. In practice, we have to work in an expansion in $k$ around $k=0$. However, 
we expect that $C$ will only get contributions at leading order in $k$ (i.e. the strict semiclassical limit), and that $B$ will be only corrected at the next-to-leading order in $k$. 
We will now verify this in ABJM theory. In contrast, the $\mu$-independent term $A$ gets corrected at all orders in $k$. 

\subsection{Quantum corrections in ABJM theory}
\label{subsect_ABJM_corrections}

We now study the general structure of quantum corrections in ABJM theory, by using the strategy explained above, i.e. by looking at the number of eigenvalues $n(E)$. Our goal 
is to show that $n(E)$ has the structure (\ref{nEst}). This involves a somewhat detailed argument. Since not every reader might go through it, we want to emphasize that the physics 
behind this argument is very simple. The WKB expansion of the density of eigenvalues of a quantum system is in fact an expansion in 
\be
\left(\hbar {\rd \over \rd E} \right)^2.
\ee
Therefore, if the leading order term in $n(E)$ is of the form $CE^2$, the first quantum correction gives the constant term in (\ref{nEst}), and further terms in the WKB expansion do not correct the polynomial part of $n(E)$. They can only give exponentially small corrections in $E$. In the rest of this section, we will verify that this qualitative argument is actually correct in the case of the one-body problem appearing in ABJM theory. 

Our starting point in the study of quantum corrections in ABJM theory is (\ref{exactdens}). As we know, there are two sources of $\hbar$ corrections in this formula. One is the quantum-corrected Hamiltonian, and the other are the terms $\CG_r$ appearing in the generalized Wigner--Kirkwood expansion. We will consider first the quantum corrections coming from $H_{\rm W}$, i.e. from the first term in (\ref{exactdens}). Since we have a symmetry $q\rightarrow -q$ and $p\rightarrow -p$ in the problem, we can restrict ourselves to the case $q\geqslant 0$ and $p\geqslant 0$. We want to solve the equation 
\be
\label{Wcurve}
H_{\rm W} (q,p)=E
\ee
in the limit $E\rightarrow \infty$. This defines a ``quantum curve" or ``quantum Fermi surface," including explicit $\hbar$ corrections. 
At leading order in $E$ the curve is given by (\ref{polygon}), and the corresponding domain (in the positive quadrant) has volume 
\be
\Vol_0(E)=2E^2.
\ee
One crucial ingredient in what follows is 
the fact that the function $U(q)$ and its derivatives have the following asymptotics as $q\gg 1$:
\be
\ba
 U(q)&=\log 2\cosh {q\over 2} ={q\over 2} +\sum_{k\geqslant1}\frac{(-1)^{k+1}}{k}\re^{-k q},\\
 U'(q)&={1\over 2} \tanh{q\over 2}={1\over 2} +\sum_{k\geqslant 1}(-1)^k \re^{-k q},\\
 U''(q)&=\frac{1}{4 \cosh^2 {q\over 2}}=\sum_{k\geqslant 1} k(-1)^{k+1} \re^{-kq}.
 \ea
 \ee
 The same results hold for $T(p)$. Notice that, if we take a number large enough of derivatives of these functions, they become exponentially suppressed at infinity. 
 This will be eventually the source of the 
 simplifications at large $E$. 

\FIGURE{
\includegraphics[height=5cm]{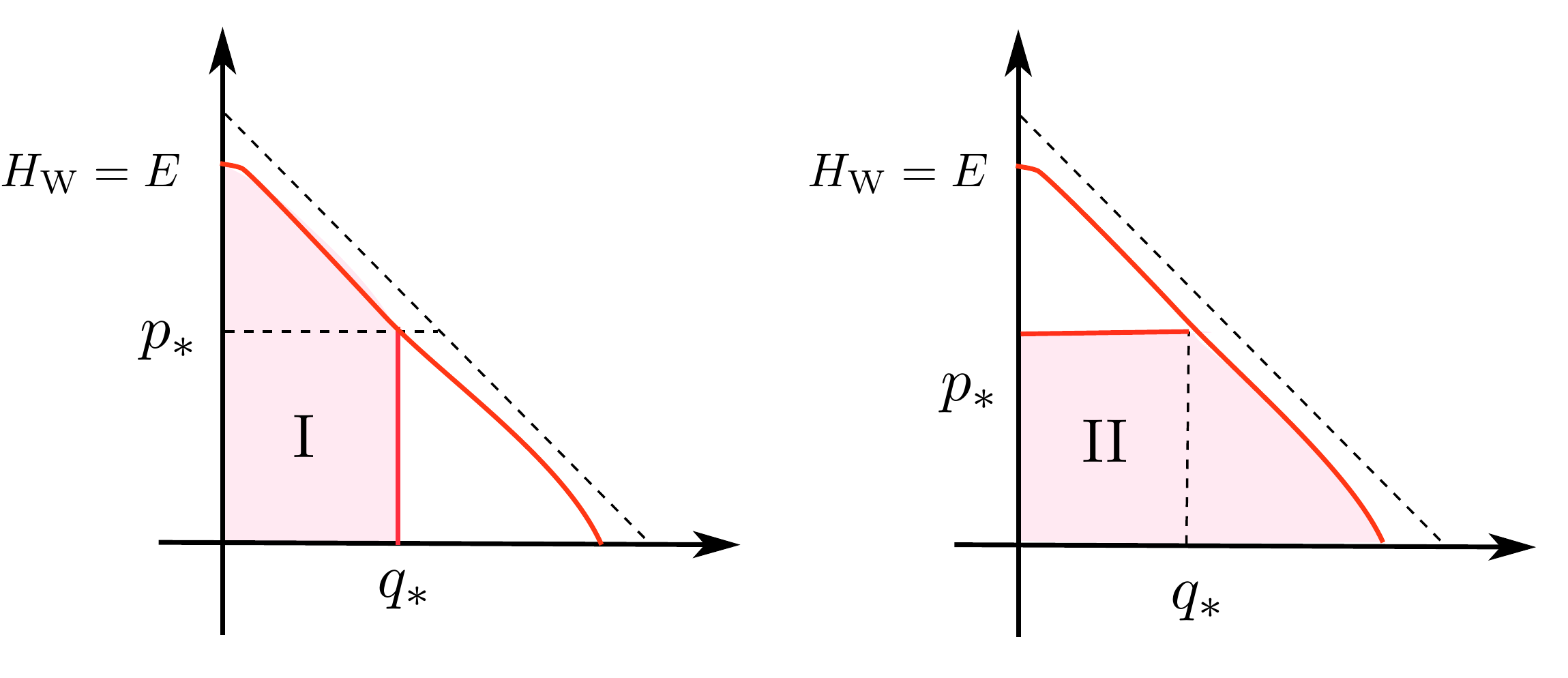} 
\caption{The regions I (left) and II (right) under the quantum curve $H_{\rm W}(q,p)=E$ in the positive quadrant. The diagonal dashed line is the polygonal curve (\ref{polygon}).}
\label{regions}
}

Let us now consider the point $(q_*, p_*)$ in the curve (\ref{Wcurve}), where
\be
p_*=E. 
\ee
It is easy to see from the explicit form of $H_{\rm W}$ that
\be
q_*=E +\CO\left(\re^{-E} \right)
\ee
where the exponentially small corrections in $E$ are themselves power series in $\hbar^2$. This point divides the curve (\ref{Wcurve}) into two segments, and 
defines two regions for the fully corrected volume, as shown in \figref{regions}. Region I is defined as the region under the quantum curve with $p\ge p_*$, 
while region II is defined by $q\ge q_*$. We have
\begin{equation}
 \Vol(E)=4\Vol_\mathrm{I}(E)+4\Vol_\mathrm{II}(E) \label{vol_I_II},
\end{equation} 
where
\begin{equation}
 \Vol_\mathrm{I}(E)=\int_0^{q_*(E)}p(E,q) \, \rd q, \qquad 
 \Vol_\mathrm{II}(E)=\int_0^{p_*(E)}q(E,p)\rd p-p_*q_*,
\end{equation} 
and $p(E,q)$ and $q(E,p)$ are local solutions of $H_{\rm W}(q,p)=E$. 

Let us first consider the curve bounding region I. Along this curve, $p(q,E)\ge E$, therefore exponential terms in $p$ in $H_{\rm W}$ 
are bounded by exponential terms in $E$. We can then write
\be
\label{firstbound}
T(p)={p\over 2}+\CO\left(\re^{-E}\right) , \qquad T'(p) = {1\over 2}+\CO\left(\re^{-E}\right), 
\ee
and
\be
\label{secondbound}
T^{(n)}(p) = \CO\left(\re^{-E}\right), \quad n\ge 2. 
\ee
In the quantum corrections to the function $H_{\rm W}$ we will have terms of the form $(T^{(k)}(p))^n$, with $k\ge 1$. Due to (\ref{firstbound}) and (\ref{secondbound}), 
and neglecting exponentially small corrections of the form $\CO\left(\re^{-E}\right)$, we should keep only the terms $(T'(p))^n$ with $n\ge1$ (like the third term in the second line of (\ref{Hw})). 
But these terms always multiply terms of the form $U^{(2n)}(q)$. We conclude that, on the curve bounding region I,  
\be
H_{\rm W}={p\over 2} +U(q)-\frac{\hbar^2}{48}U''(q)+\frac{1}{2}\sum_{n>1}\hbar^{2n} c_n U^{(2n)}(q) +\CO\left(\re^{-E}\right).
\label{h_exp_I}
\ee
The third term in this expression comes from the third term in the second line of (\ref{Hw}). The fourth term comes from higher quantum corrections (see the first term in the 
last line of (\ref{Hwtwo}) for an example of such a term at order $\CO(\hbar^4)$). We can now solve for $p$ along this curve, 
\begin{equation}
 p(E,q)=2E-q+\Delta p(E,q),
\end{equation} 
where
\begin{equation}
 \Delta p(E,q)=q-2U(q)+\frac{\hbar^2}{24}U''(q)-\sum_{n>1}\hbar^{2n} c_n U^{(2n)}(q) +\CO(\re^{- E}).
\end{equation}
We calculate the volume of region I as follows, 
\begin{equation}
 \Vol_\mathrm{I}=\Vol^{0}_\mathrm{I}+\Delta\Vol_\mathrm{I}.
\end{equation} 
The first term comes from the polygonal limit of the curve, 
\begin{equation}
 \Vol^0_\mathrm{I}(E)=\int\limits_0^{q_*(E)}(2E-q) \rd q=2 E q_*(E)-{q_*^2(E) \over 2}.
\end{equation} 
The second term comes from the corrections to the curve, and it is given by  
\be
\ba
 \Delta\Vol_\mathrm{I}(E)&=\int\limits_0^{q_*(E)}\Delta p(E,q) \rd q\\
& -2 \int\limits_0^{q_*(E)} \left( U(q)-{q \over 2} \right) \rd q+\frac{\hbar^2}{24}\int\limits_0^{q_*(E)} U''(q) \rd q-\sum_{n>1}\hbar^{2n} c_n \int\limits_0^{q_*(E)}U^{(2n)}(q)+
\CO\left( E \re^{-E} \right)\\
&=-2 \int\limits_0^{\infty} \left( U(q)-{q \over 2} \right) \rd q+\frac{\hbar^2}{24}\int\limits_0^{\infty} U''(q) \rd q-\sum_{n>1}\hbar^{2n} c_n \int\limits_0^{\infty}U^{(2n)}(q)+
\CO\left( E \re^{-E} \right)
\\
&= -\frac{\pi^2}{6}+\frac{\hbar^2}{48}+\CO\left( E \re^{-E} \right).
\ea
\ee
In the last calculation we used that, up to non-perturbative terms in $E$, we can extend the integration region to infinity, and also that
\begin{equation}
  \int\limits_0^\infty U^{(2n)}(q) \rd q=U^{(2n-1)}(\infty)-U^{(2n-1)}(0)=0 \quad \text{for $n>1$.}
\end{equation}
A similar calculation can be done for region II. We obtain, from the polygonal approximation of the curve, 
\begin{equation}
 \Vol^0_\mathrm{II}(E)=2Ep_*(E)-\frac{p_*^2(E)}{2}-p_*(E)q_*(E),
\end{equation} 
while the corrections give, 
\begin{equation}
 \Delta\Vol_\mathrm{II}(E)=-\frac{\pi^2}{6}-\frac{\hbar^2}{96}+\CO\left( E \re^{-E} \right).
\end{equation} 
Using that
\begin{equation}
 p_*(E)+q_*(E)=2E+\CO\left( \re^{-E} \right)
\end{equation} 
we finally get
\begin{equation}
 \Vol(E)=
{8E^2}-\frac{4\pi^2}{3}+\frac{\hbar^2}{24}+\CO\left( E \re^{-E} \right).
\end{equation} 

We now consider the contribution from the quantum corrections to the density. In fact, these terms only give non-perturbative corrections in $E$. 
Using that
\begin{equation}
 \delta(E-H_{\rm W}(q,p))=\delta(p-p(E,q))\left/\frac{\d H_{\rm W}(q,p)}{\d p}\right.
 =\delta(q-q(E,p))\left/\frac{\d H_{\rm W}(q,p)}{\d q}\right.
\end{equation} 
one can always decompose an integral over the phase space as a sum of one-dimensional integrals in regions I and II, as in (\ref{vol_I_II}). 
For region I one can use again the expression (\ref{h_exp_I}) and the properties (\ref{firstbound}), (\ref{secondbound}). The only nontrivial term which gives 
an $\hbar^2$ correction comes from $\CG_3$ and gives, 
\begin{multline}
 \frac{\hbar^2}{24}\frac{\d^2}{\d E^2}\int\limits_0^{q_*(E)} \rd q 
 \left.\frac{\d H_{\rm W}(q,p)}{\d p}\frac{\d^2 H_{\rm W}(q,p)}{\d q^2}\right|_{p=p(E,q)}=\\
 \frac{\hbar^2}{24}\frac{\d^2}{\d E^2}\int\limits_0^{q_*(E)} \rd q \left(\sum_{n\geq 0}\hbar^{2n} c_n U^{(2n+2)}(q)+\CO(\re^{-E})\right)=\CO(E \re^{- E}).
\end{multline}
For higher order corrections everything that contains a term with 
\be
{\d^r H_{\rm W}(q,p) \over \d p^r}, \qquad r>1, 
\ee
or with 
\be
{\d^2 H_{\rm W}(q,p) \over \d p\d q}
\ee
is of order $\re^{-E}$. Since the derivatives $\partial_p$ and $\partial_q$ always come in pairs, the only terms possibly contributing are of the form
\begin{equation}
 \prod_i\left(\frac{\d H_{\rm W}(q,p)}{\d p}\right)^{n_i}\frac{\d^{n_i} H_{\rm W}(q,p)}{\d q^{n_i}}=
\prod_i\frac{\d^{n_i} H_{\rm W}(q,p)}{\d q^{n_i}}+\CO(\re^{-E})
\end{equation} 
where $n_i\geq 2,\forall i$. After integrating and applying $\d^r/\d E^r$ this gives a correction of order $\CO(E \re^{-E})$ by the same reason. 

We conclude that, to all orders in the $\hbar$ expansion, 
\begin{equation}
\label{all-density}
n(E)=\frac{\Vol(E)}{2\pi\hbar}+\CO(E \re^{- E})=
\frac{2E^2}{\pi k}-\frac{1}{3k}+\frac{k}{24}+\CO(E \re^{- E}).
\end{equation} 
Therefore, by using (\ref{Jden}) and (\ref{shift_rel}) we find the expression
\be
 J_{\rm ABJM}(\mu) ={2 \mu^3 \over 3 k \pi^2} + \mu  \left( {1\over 3k} +{k \over 24}\right) +A(k)+J_{\rm np}(\mu)
 \ee
where 
\be
J_{\rm np}(\mu)=\sum_{\ell,n=1}^{\infty} \left(a_{\ell,n} \mu^2 + b_{\ell,n} \mu + c_{\ell,n} \right) k^{2n-3} \re^{-2 \ell \mu}.
\ee
It is not manifest from the above results that this series involves only even powers of $z^{-1}$, but we have verified it to be the case for the first three orders 
in $k$, and we believe it is a general feature. Finally, it follows from (\ref{all-density}) and (\ref{airydens}) that 
\be
\label{abjm-airy}
 Z_{\rm ABJM}(N)=C^{-1/3}\re^{A(k)}\Ai\left[C^{-1/3}\left(N-{1\over 3k} -{k \over 24} \right)\right] + Z_{\rm np}(N), 
 \ee
 where $C$ is given in (\ref{Ck}) and $Z_{\rm np}(N)$ are exponentially suppressed corrections at large $N$. This concludes our derivation of the Airy behavior 
 for ABJM theory. The function $A(k)$ can in principle be determined, order by order in $k$, by computing the $Z_\ell^{(n)}$, resumming the resulting series, and expanding at $\mu=\infty$, as we did in sections \ref{largencorabjm} and \ref{firstqc}. One obtains, 
 \be
 \label{aexp}
 A(k)={2 \zeta(3) \over \pi^2 k} -{k \over 12} - {\pi^2 k^3 \over 4320} +\CO(k^5). 
 \ee
A sketch of the computation leading to the third term of this expansion can be found in the Appendix. In principle one can also compute $A(k)$ by using the representation (\ref{shift_rel}). What is the interpretation of $A(k)$? One natural possibility is that $A(k)$ encodes effects of order
\be
\label{D0}
\CO\left(\re^{-k} \right) \sim  \CO\left(\re^{-1/g_s}\right), 
\ee
i.e. that $A(k)$ gives the contribution from D0 branes. Notice that the second and third coefficients in (\ref{aexp}) are given by 
\be
\label{ansAk}
-{|B_{2g}| \over g (2g-1) (2g)!} \pi^{2g-2} 
\ee
for $g=1,2$. These are the coefficients of the power series expansion in $k$ of 
\be
-{2\over \pi} \int_0^{\pi k} {\rd \xi \over \xi^2} \log\left[ { \sin \left(\xi /2\right) \over  \xi /2 }\right].
\ee
This gives indeed corrections of the form (\ref{D0}). It would be interesting to verify that the above function resums the small $k$ expansion of $A(k)$\footnote{After the first version of this paper 
was submitted, numerical and analytical studies of the function $A(k)$ were performed in \cite{hanada}. The ansatz (\ref{ansAk}) proposed here turns out to be incorrect. 
The results of \cite{hanada} strongly suggest that the function $A(k)$ can be obtained by resumming the so-called constant map contribution to the free energy, and expanding it around $k=0$. The resulting series reproduces (\ref{aexp}). We would like to thank the authors of \cite{hanada} for informing us of their result prior to publication, which prompted us to correct a minor sign error in the calculation of (\ref{aexp}).}. 

\subsection{Quantum-mechanical instantons as worldsheet instantons}
\label{qmisec}

One obvious question that one can ask at this point is the following: 
where are the worldsheet instantons (\ref{winst}) that one finds perturbatively in $g_s$ in the 't Hooft expansion? 
We now give some preliminary evidence that worldsheet instantons correspond to the quantum-mechanical instantons of the Hamiltonian $H_{\rm W}$. 

So far our focus has been in the perturbative corrections in $\hbar$, but one should expect generically non-perturbative corrections due to instantons, 
of order $\exp(-1/\hbar)$. To understand these quantum-mechanical instantons in our problem, with a non-conventional Hamiltonian, 
we need a general, geometric approach to non-perturbative WKB expansions, like the one 
proposed in \cite{vorosquartic,ddp}. In this approach, instanton contributions are obtained by looking at the complexified curve
\be
H(q,p)=E
\ee
where $H(q,p)$ is the Hamiltonian of the model. 
Perturbative WKB expansions are associated to periods of the above curve around ``A-type" cycles, while non-perturbative corrections to the WKB method are 
associated to ``B-type" cycles. 
\begin{figure}
\begin{center}
\includegraphics[scale=.7]{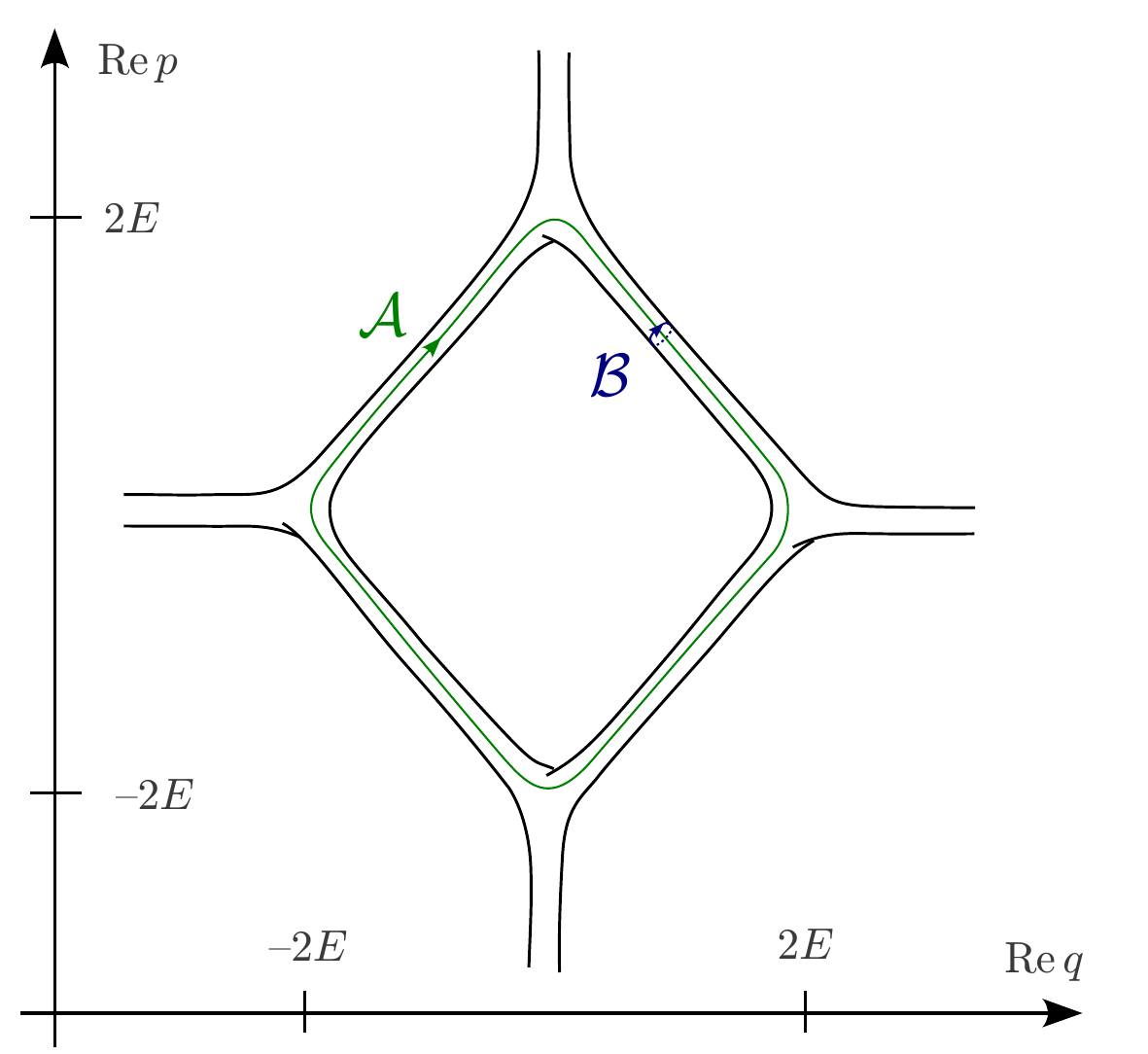}
\end{center}
\caption{The Riemann surface of $H_\mathrm{cl}(p,q)=E$ for ABJM theory for large $E$. The four interior tubes form the limiting polygon of the Fermi surface.} 
\label{fig_fermirs}
\end{figure}
In the case of ABJM theory, the complexified curve is identical to the spectral curve (\ref{sc}), after an appropriate choice of the variables. Its Riemann surface looks as shown in the Fig.~\ref{fig_fermirs}. Let us introduce canonical coordinates $Q$, $P$ related to the $q$, $p$ coordinates as 
\be
Q=q, \qquad P=p+q. 
\ee
This preserves the symplectic form. The coordinate $P$ is chosen so that it has no monodromy along the contour $\mathcal{B}$. Then in the large $E$ limit
\begin{equation}
 n(E)\approx \frac{1}{2\pi\hbar}\oint_\mathcal{A}P\rd Q\approx {1\over 2\pi\hbar} \Vol\left\{(q,p): |p|+|q|<2E\right\}=
 { 4 E^2 \over \pi \hbar}.
\end{equation}
The instanton contribution is of order
\begin{equation}
 \exp\left[ {\frac{\ri}{\hbar}\oint_\mathcal{B}P\rd Q} \right],
\end{equation} 
where in the large $E$ limit 
\begin{equation}
 \oint_\mathcal{B}P\rd Q= 2E\cdot 4\pi\ri+\mathcal{O}(\re^{-cE}). \label{instanton_action_behavior}
\end{equation}
Here we used that, in the interior of the upper-right tube, 
$P=p+q=2E+\mathcal{O}(\re^{-cE})$ for some constant $c$, and that the monodromy of $Q$ around the tube is $4\pi\ri$. The above period can be computed exactly with the results of \cite{dmpnp} since the behavior (\ref{instanton_action_behavior}) fixes it completely:
\begin{equation}
\label{Bper}
\ba
 \oint_\mathcal{B} P\rd Q& =-2\ri \re^{E} \pi  \, _3F_2\left(\frac{1}{2},\frac{1}{2},\frac{1}{2};1,\frac{3}{2};\frac{\re^{2E}}{16}\right)-\frac{\re^{E}}{\pi} G_{3,3}^{2,3}\left(\frac{\re^{2E}}{16}\left|
\begin{array}{c}
 \frac{1}{2},\frac{1}{2},\frac{1}{2} \\
 0,0,-\frac{1}{2}
\end{array}
\right.\right)+4 \pi ^2\\
&=8\ri\pi E+\mathcal{O}(\re^{-2E}).
\ea
\end{equation} 
In fact, after the identification (\ref{kappa_energy}), this period is equal to $-4A_s$, where $A_s$ is the strong coupling instanton action computed in \cite{dmpnp}. For large energy, (\ref{Bper}) gives a contribution to the density of states of order
\begin{equation}
 \exp\left[-4E/k\right]
\end{equation}
which becomes a contribution 
\begin{equation}
\label{winstmu}
\exp\left[-4\mu/k \right] 
\end{equation}
to the grand canonical potential, and a contribution 
\be
\sim \exp\left[ -2\pi\sqrt{2N/k}\right] 
\ee
to the canonical free energy. This is precisely the weight of a worldsheet instanton (\ref{winst}) in ABJM theory. 

Quantum-mechanical instantons are of course invisible in the perturbative $\hbar$ expansion of $H_{\rm W}$ and in the Wigner--Kirkwood expansion, but they appear in the 
't Hooft expansion. In fact, the 't Hooft expansion of the canonical free energy 
\be
F(\lambda, k)= \sum_{g\ge 0} k^{2-2g} F_g(\lambda) 
\ee
 leads to a genus expansion of the grand canonical potential of the form \cite{kkn}
 \be
 \label{Jthooft}
 J_{\text{ 't Hooft}}(\mu, k)=\sum_{g\ge 0} k^{2-2g} \CJ_g(\mu/k). 
 \ee
Notice that, in the Fermi gas approach, only the perturbative part in $\mu$ of $J(\mu)$ can be written in this form (\ref{Jthooft}). 
The membrane instanton contributions and the function $A(k)$ do not have the right 
functional dependence in $\mu/k$ to fit into the 't Hooft expansion, while the weight associated to a quantum-mechanical instanton (\ref{winstmu}) is again of the right form. 
In the case of ABJM theory, we see from (\ref{JABJM}) that the Fermi gas approach gives 
\be
\label{topJ}
\ba
\CJ_0(\zeta)&={2 \zeta^3 \over 3 \pi^2} + {\zeta \over 24}+\CO\left(\re^{-4 \zeta}  \right), \\
\CJ_1(\zeta)&={\zeta \over 3}+\CO\left(\re^{-4 \zeta } \right),\\
\CJ_g(\zeta)&=\CO\left(\re^{-4 \zeta  } \right), \qquad g\ge 2, 
\ea
\ee
where $\zeta=\mu/k$. From the point of view of the topological string, it follows from (\ref{Tmu}) that the variable $\zeta$ is essentially the period $T$ at large radius, and a perturbative Fermi gas approach makes possible to recover the leading, perturbative genus zero and genus one free energies of the topological string given in (\ref{LRfree}). 
 
Finally, we should mention that there is an extra source of worldsheet instanton-like corrections. In general, the exact representation (\ref{exactinverse}) and the saddle-point integral (\ref{muint}) are only equivalent up to exponentially small corrections in $N$. Since we are taking into account such corrections, we have to be more careful here. The expression (\ref{exactinverse}) is equivalent to 
\be 
\label{exactmu}
Z(N)={1\over 2 \pi \ri} \int_{\mu_*-\ri \pi}^{\mu_*+\ri \pi} \rd \mu \, \exp \left[ J(\mu) -\mu N\right], 
\ee
where the integration contour is parallel to the imaginary axis, and $\mu_*$ is arbitrary. To apply the saddle-point method, one chooses for $\mu_*$ the saddle-point 
of the exponent, and then extends the integration contour to infinity along the imaginary axis 
(this is what gives the Airy function behavior we have found many times in this paper). As it is well-known, it is in this last step of extending the integration contour 
that one introduces exponentially small errors in $N$. A rough estimate of these errors can be done 
as follows. The saddle-point expansion involves integrating a Gaussian of the form 
\be
 \exp\left[{1\over 2} (\mu-\mu_*)^2 J''(\mu_*) \right].
\ee
The error in going to (\ref{muint}) can then be estimated by evaluating this Gaussian at the true endpoints in (\ref{exactmu}). This gives, by using the leading term in (\ref{JABJM}),
\be
\sim \exp\left[-2\mu_*/k \right] 
\ee
which is the square root of (\ref{winstmu}). Therefore, these type of corrections should also be taken in account when trying to extract information about worldsheet instantons. 

\sectiono{More general Chern--Simons--matter theories}
\label{genCSM}

In this section we consider in detail more general CSM theories. We first study the thermodynamic limit of necklace quivers, and derive a general formula 
for the large $N$ limit of their free energy which agrees with the result obtained in \cite{herzog,herzog2} by analyzing the matrix model. Then we extend the considerations of section \ref{subsect_ABJM_corrections} to the general necklace CSM theories considered in section \ref{subsect_generalCSM}. 
For technical reasons we restrict ourselves to theories 
whose Hamiltonian is Hermitian, i.e. whose free energy is real, and we show that, with that assumption, the Airy behavior of the resummed $1/N$ expansion 
found in \cite{fhm} is indeed generic. 
These general considerations are then illustrated in detail in the case of the ABJM theory (i.e. the two-node theory) with fundamental matter. Finally, we consider the massive theories of \cite{gt}, and we derive the $N^{5/3}$ behavior found in \cite{ajtz,jkps} with our techniques. 

\subsection{Thermodynamic limit for general necklace quivers}
In this section we study the general necklace quiver considered in subsection \ref{subsect_generalCSM}. 
From (\ref{rho_wigner_quiv}) it follows that, for large energy,
the Fermi surface is defined by the polygonal equation 
\be
\label{nodepolygon}
\sum_{j=1}^r\left|p-\left(\sum_{i=1}^{j-1}n_i\right) q\right|+\left(\sum_{j=1}^r\frac{N_{f_j}}{k}\right)|q|=2E. 
\ee
\FIGURE{
\includegraphics[height=6cm]{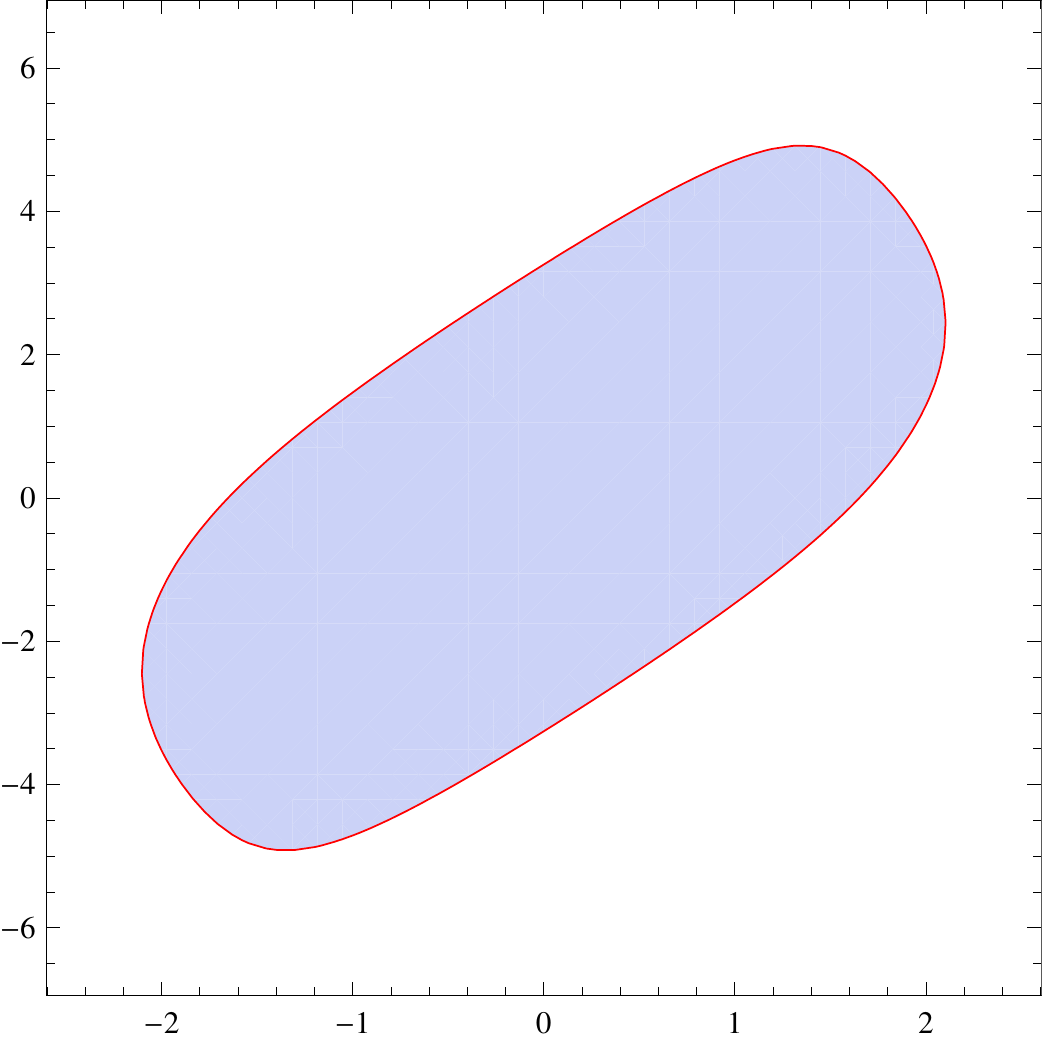} \qquad 
\includegraphics[height=6cm]{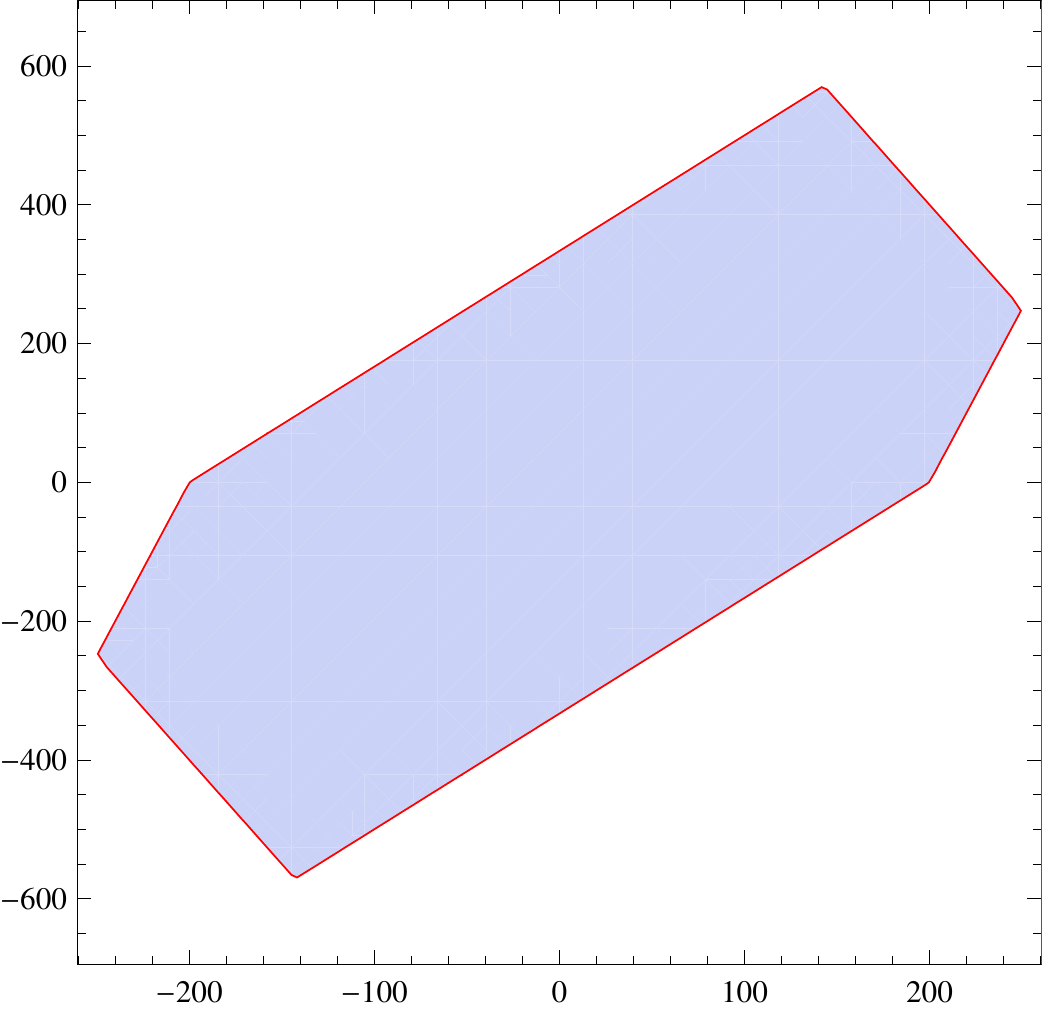}
\caption{The Fermi surface for the three-node quiver with $n_a=(1,3,-4)$ in the $q$-$p$ plane, for $E=5$ (left) and $E=500$ (right). At large energy it  
approaches the polygon (\ref{nodepolygon}).}
\label{threenode}
}
As an example, we show in \figref{threenode} the classical Fermi surface (\ref{nodepolygon}) at small and large energy, for a three-node quiver with $n_a=(1,3,-4)$. By the by now familiar argument of the previous sections, the number of eigenstates is given by the semiclassical formula (\ref{semivol}) applied 
to the domain bounded by (\ref{nodepolygon}). One finds, 
\be
n(E) \approx CE^2, 
\ee
where the constant $C$ is given by 
\be
\pi^2 C=\,\Vol\left\{(x,y)\, : \,\sum_{j=1}^r\left|y-\left(\sum_{i=1}^{j-1}k_i\right) x\right|+\left(\sum_{j=1}^r N_{f_j}\right)|x|< 1\right\}
 \label{necklace_volume}
\ee
and the variables $y,x$ differ from $p,q$ in (\ref{nodepolygon}) by rescaling. Once this constant has been determined, the large $N$ asymptotics of the free energy 
is given by (\ref{freeenergy})
\begin{equation}
 F(N)\approx -\frac{2}{3}C^{-1/2}N^{3/2}.
\end{equation} 
In order to compute $C$, we notice that the right hand side of (\ref{necklace_volume}) is the volume
 of a convex polygon which can be easily calculated. Suppose for simplicity that $N_{f_j}=0$. Let us introduce new parameters $c_j$, 
 related to $k_i$ in the following way
\begin{equation}
 c_{\sigma(j)}=c+\sum_{i=1}^{j-1}k_i
\end{equation} 
where $c$ is an auxiliary constant and $\sigma$ is a permutation chosen so that $c_i\leq c_{i+1},\forall i$. Then
\begin{equation}
 \pi^2 C=\Vol\left\{(x,y)\,: \,\sum_{j=1}^r \left|y-c_j x\right|< 1\right\}. 
\label{volume_cs}
\end{equation}
Let us note that the $c_j$ differ by a permutation from the parameters $q_a$ introduced in \cite{hklebanov} (where they were defined in such a way 
that $k_a=q_{a+1}-q_a$). Notice also that the expression (\ref{volume_cs}) is explicitly invariant under permutations of the $c_j$. 
The inequality $\sum_{j=1}^r\left|y-c_j x\right|< 1$ defines a convex hull of $2r$ points $(\pm x_s,\pm y_s)$ so that
\be
y_s=c_sx_s, \qquad \sum_{j=1}^r \left|y_s-c_j x_s\right|=1.
\ee
One finds
\begin{equation}
 x_s=\frac{1}{\sum_{j=1}^r \left|c_s-c_j\right|},\qquad y_s=\frac{c_s}{\sum_{j=1}^r \left|c_s-c_j\right|}.
\end{equation}
Then one can use the standard formula for the area of a convex hull to find
\begin{equation}
\label{chull}
 \pi^2 C=\sum_{s=1}^{r}\frac{|c_{s+1}-c_s|}
 {\left(\sum_{j=1}^r \left|c_{s+1}-c_j\right|\right)\left(\sum_{j=1}^r \left|c_s-c_j\right|\right)}
\end{equation} 
where as usual we use the convention $c_{r+1}\equiv c_1$. Let us illustrate this formula by applying it to necklaces with three and four nodes. 
For the necklace with three nodes $(k_1,k_2,k_3)$ we can assume, without loss of generality, that 
\be
|c_1-c_2|=|k_3|, \qquad |c_2-c_3|=|k_1|, \qquad |c_1-c_3|=|k_2|.
\ee
Then
\begin{equation}
 \frac{\pi^2 C}{2}=\,\frac{|k_1||k_2|+|k_2||k_3|+|k_3||k_1|}{(|k_1|+|k_2|)(|k_2|+|k_3|)(|k_3|+|k_1|)}.
\end{equation}
For the quiver with four nodes, 
let us assume without loss of generality that $\sum_{i=1}^4{c_i}=0$. Then an easy computation gives
\begin{equation}
 \frac{\pi^2 C}{2}=\frac{1}{32}\left(\frac{1}{c_1}-\frac{1}{c_4}+12\,\frac{1}{c_3+c_4}+4\,\frac{c_1+c_3}{(c_3+c_4)^2}\right).
\end{equation}

These formulae for the three and four-node quivers agree with the results first found in \cite{hklebanov} (for the four-node quiver, their formula is 
obtained by setting $c_1=q_3$, $c_2=q_1$, $c_3=q_2$, $c_4=q_4$). In fact, the above general result for the free energy of these CSM theories, 
involving the area of the polygon (\ref{necklace_volume}), has been derived in this form in \cite{herzog,herzog2} by refining the analysis of the matrix model done in 
\cite{hklebanov} (where a different, but equivalent general formula for the free energy 
was proposed). This class of CSM theories is dual to M-theory on $\mathrm{AdS}_4\times X_7$, where $X_7$ is an appropriate 
tri-Sasaki Einstein space \cite{quiver2}. Therefore, the coefficient $C$ should be proportional to the volume of the $X_7$ manifold, and 
one should have
\begin{equation}
 \frac{\Vol(X_7)}{\Vol(\IS^7)}=\frac{\pi^2 C}{2}=\frac{1}{2}\,\Vol\left\{(x,y)\, : \,\sum_{j=1}^p\left|y-\left(\sum_{i=1}^{j-1}k_i\right) x\right|+\left(\sum_{j=1}^pN_{f_j}\right)|x|< 1\right\}.
 \label{necklace_quotient}
\end{equation}
This is indeed the case, as it was proved in \cite{herzog}.

We then see that the Fermi gas approach allows us to rederive the result for the large $N$ energy obtained in \cite{herzog}, but in a simpler way. The polygon 
appearing in the matrix model analysis of \cite{herzog} has here a very simple interpretation: it is the large energy limit of the Fermi surface for the ideal Fermi gas. 

\subsection{Airy function behavior for a class of CSM theories}
\label{airy-general}
We now extend the considerations of section \ref{subsect_ABJM_corrections} to more general necklace quivers with matter. In the next subsection we apply the 
general considerations developed here to the case of ABJM theory with matter.

Let us assume that the Wigner transform of the density matrix can be written in the form
\begin{equation}
 \rho_\wigner(q,p)\equiv \exp_\star\left\{-H_W\right\}=\re^{-\Phi_1(Q_{R_1})}\star \re^{-\Phi_2(Q_{R_2})}\star\cdots\star\re^{-\Phi_m(Q_{R_m})},
 \label{airy_KW}
\end{equation}
where 
\be
Q_R=a_Rq+b_Rp
\ee
for suitable $a_R$, $b_R$, and the different $Q_{R}$ are given by linearly independent combinations of $q, p$. 
We will also suppose that the functions $\Phi_i$ are real valued, even, and that\footnote{In what follows, the constant $c>0$ may have different values in different formulae.}
\begin{equation}
 \Phi_i(Q)=\gamma_i|Q|+\mathcal{O}(\re^{-c|Q|}),\qquad Q\rightarrow\infty,\quad \gamma_i>0.
\end{equation}
These assumptions are obviously true for the general necklace quivers considered in subsection \ref{subsect_generalCSM}. In addition, we will 
suppose that $H_\wigner(q,p)$ is real. This corresponds to the case when the quantum operator $\hat{H}$ is Hermitian. 
It should be possible to treat the general case, when $\hat{H}$ has complex eigenvalues, with similar techniques, but we will not do it here. 

As usual, the leading contribution to the number of states is given by the volume of a polygon:
\begin{equation}
 n(E)\approx{1\over 2\pi\hbar} \Vol \left\{ (q,p): \sum_{i=1}^m\gamma_i|a_{R_i}q+b_{R_i}p|<E \right\}=CE^2.
\end{equation} 
\begin{figure}
\center
 \includegraphics{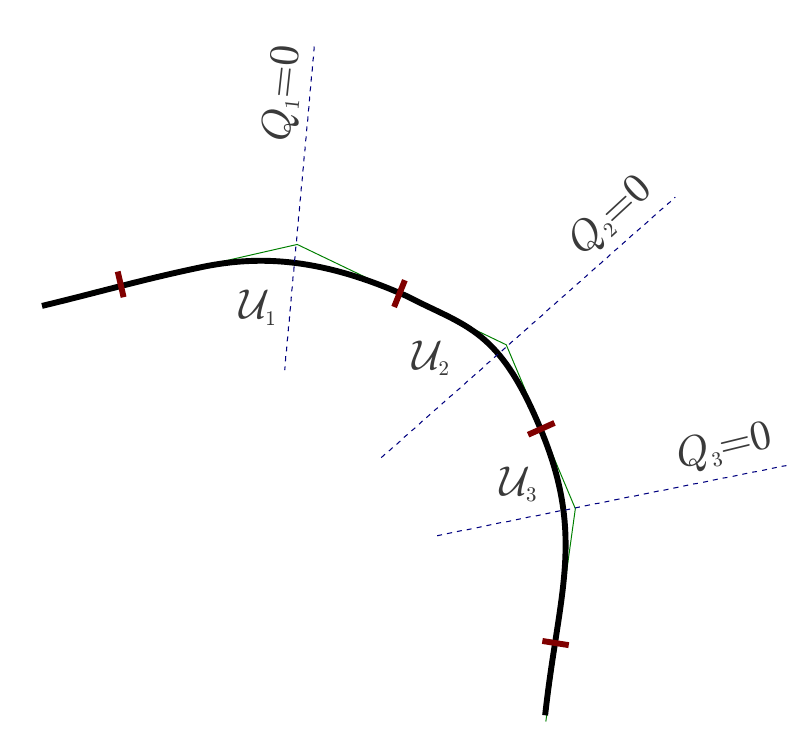}
\caption{The black thick line depicts the Fermi surface (\ref{fermi-curve}). The green thin line depicts the limiting polygon. 
The red thick dashes mark the boundaries of the patches $\mathcal{U}_R$.}
\label{general_fermi_surface}
\end{figure}
Let 
\be
\label{fermi-curve}
\mathcal{C}_E=\left\{(q,p)\,:\,H_\wigner(p,q)=E\right\}
\ee
be the real curve describing the Fermi surface. One can always decompose $\mathcal{C}_E$ into patches $\mathcal{U}_R$, $\mathcal{U}_R'$ so that both 
$\mathcal{U}_R$ and $\mathcal{U}_R'$ contain one point where $Q_R=0$, $\mathcal{U}_R'$ is related by to $\mathcal{U}_R$ by reflection through the center 
of the polygon, and
\begin{equation}
 \left.\sum_{i=1}^m\gamma_i|a_{R_i}q+b_{R_i}p|\right|_{\d\mathcal{U}_R,\,\d\mathcal{U}_R'}=E+\mathcal{O}(\re^{-cE}).
 \label{boundary_estimate}
\end{equation}
In \figref{general_fermi_surface} we depict the general structure of such decomposition. In the case of ABJM theory considered in section \ref{subsect_ABJM_corrections}, the boundaries of the regions I and II lying on the curve $\mathcal{C}_E$ are halves of the patches $\mathcal{U}_1$ and $\mathcal{U}_2$. A particular example of such a decomposition of the Fermi surface in regions is shown in \figref{matter_regions}, in the case of ABJM theory with fundamental matter. We will now argue that, for each patch $\mathcal{U}_R$, the structure of 
corrections is essentially the same as in the ABJM case.

One can always choose 
\be
P_R=c_Rp+d_Rq
\ee
so that 
\be
\rd Q_R\wedge \rd P_R=\rd q\wedge \rd p 
\ee
and such that, in the domain $\mathcal{U}_R$, the Hamiltonian can be written as follows:
\begin{equation}
 H_\wigner(q,p)=\beta_R P_R+\sum_{i|R_i=R} \Phi_i(Q_R)+\sum_{r>0}\sum_{i|R_i=R}\hbar^{2r} c^R_{r,i}\Phi^{(2r)}_i(Q_R)+\mathcal{O}(\re^{-cE})
\end{equation}
where $\mathcal{O}(\re^{-cE})$ denotes an estimate which is uniform in $\mathcal{U}_R$. Without loss of generality one can assume that $\beta_R>0$. Then in the domain $\mathcal{U}_R$ the solution to $H^{(R)}_W(P_R,Q_R)\equiv H_W(p,q)=E$ can be written as
\begin{equation}
 P_R(E,Q_R)=\frac{1}{\beta_R}\left(E-\sum_{i|R_i=R} \gamma_i|Q_R|\right)+\Delta P_R(E,Q_R). 
\end{equation} 
In this equation, 
\begin{equation}
 \Delta P_R(E,Q_R)=\Delta_\rp P_R(Q_R)+\mathcal{O}(\re^{-cE}), 
 \label{local_solution_estimate}
\end{equation} 
where $\Delta_\rp$ denotes the perturbative part of the correction. It is given by 
\begin{equation}
 \Delta_\rp P_R(Q_R)=-\frac{1}{\beta_R}\left(\sum_{i|R_i=R} (\Phi_i(Q_R)-\gamma_i|Q_R|)+\sum_{r>0}\sum_{i|R_i=R}\hbar^{2r} c^R_{r,i}\Phi^{(2r)}_i(Q_R)\right),
\end{equation}
and satisfies the property,
\begin{equation}
 \Delta_\rp P_R(Q_R)=\mathcal{O}\left(\re^{-c|Q_R|}\right),\; Q_R\rightarrow\infty
 \label{correction_estimate}
\end{equation} 
The property (\ref{boundary_estimate}) implies that
\begin{equation}
 \Vol\left\{H_\wigner(q,p)<E\right\}=\Vol\left\{\sum_{i=1}^m\gamma_i|Q_R|<E\right\}+2\sum_R\Delta\Vol_R(E)+\mathcal{O}(\re^{-cE})
\end{equation}
where
\begin{equation}
 \Delta\Vol_R(E)=\int\limits_{\mathcal{U}_R}\Delta  P_R(E,Q_R)\rd Q_R.
\end{equation}
From (\ref{local_solution_estimate}) it follows that
\begin{equation}
 \Delta\Vol_R(E)=\int\limits_{\mathcal{U}_R}\Delta_\rp P_R(Q_R)\rd Q_R+\mathcal{O}(\re^{-cE}).
\end{equation}
By using (\ref{correction_estimate}), we can extend the integration region to infinity, up to non-perturbative corrections, and we obtain
\begin{equation}
 \Delta\Vol_R(E)=\int\limits_{-\infty}^\infty\Delta_\rp P_R(Q_R)\rd Q_R+\mathcal{O}(\re^{-cE}).
\end{equation}
Let us denote
\begin{equation}
 \Delta_\rp\Vol_R \equiv \int\limits_{-\infty}^\infty\Delta_\rp  P_R(Q_R)\rd Q_R =
 -\frac{1}{\beta_R}\int\limits_{-\infty}^\infty\sum_{i|R_i=R}(\Phi_i(Q_R)-\gamma_i|Q_R|)\rd Q_R
- \frac{2\hbar^{2}}{\beta_R}\sum_{i|R_i=R} c^R_{1,i}\gamma_i.
\end{equation}
Similarly to what happened in ABJM theory, there are no perturbative corrections to $n(E)$ from higher terms of the Wigner--Kirkwood expansion. Therefore we obtain
\begin{equation}
 n(E)=CE^2+n_0+\mathcal{O}(\re^{-cE})
\end{equation} 
with
\begin{equation}
\label{n0alg}
 n_0=\frac{1}{\pi\hbar}\sum_{R}\Delta_\rp\Vol_R.
\end{equation}
As was shown in subsection \ref{subsect_generalcorr}, it then follows that the $1/N$ corrections are resummed to an Airy function
\begin{equation}
 Z(N)=C^{-1/3}\re^{A}\Ai\left[C^{-1/3}(N-B)\right]+Z_\np(N)
\end{equation}
where
\begin{equation}
 B=n_0+\frac{C\pi^2}{3}.
\end{equation}
The above general argument gives an explicit algorithm to compute the constant (\ref{n0alg}). 
In the next subsection we consider the example of the ABJM theory with matter as an illustration of this argument. 

\subsection{ABJM theory with fundamental matter}

\begin{figure}
\center
 \includegraphics[scale=1.5]{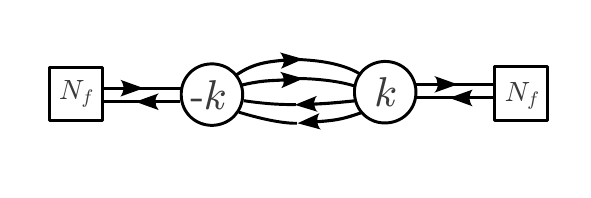} 
\caption{Quiver for the two-node theory with fundamental matter.}
\label{general_necklace}
\end{figure}

The ABJM theory with matter we will consider is described by a two-node quiver with equal number of fundamentals in each node, see \figref{general_necklace}. 
The density matrix of this theory is given by (\ref{rho_wigner_matter}):
\begin{equation}
 \rho_\wigner\equiv \exp_\star\{- H_\wigner\}=
\re^{-U(q)/2}\star \re^{-\Psi(p+q)} \star \re^{-T(p)}\star \re^{-\Psi(p+q)} \star \re^{-U(q)/2} \label{matter_KW}
\end{equation} 
where $U(q)$, $T(p)$ are given respectively in (\ref{upotential}) and (\ref{kinabjm}), and 
\begin{equation}
 \Psi(p+q)=N_f\log 2\cosh\frac{p+q}{2k}.
\end{equation}
The Wigner transform of the Hamiltonian has the following $\hbar$ expansion, which can be obtained with the use of the Baker-Campbell-Hausdorff formula:
\be
\ba
& H_\wigner(p,q)=H_\mathrm{cl}(p,q)+\frac{\hbar ^2}{24}  U'(q)^2 \left(2 \Psi ''(p+q)+T''(p) \right)\\ 
&-{\hbar^2 \over 12} \left(-2 \Psi '(p+q)^2 \left(T''(p)-2 U''(q)\right)+T'(p)^2 \left(2 \Psi ''(p+q)+U''(q)\right)+4 T'(p) U''(q) \Psi '(p+q)\right)\\
&+{\hbar^2 \over 6} U'(q) \left(T''(p) \Psi '(p+q)-T'(p) \Psi ''(p+q)\right)+\CO\left(\hbar ^4\right),
\ea
\ee
where the first term is the ``classical'' Hamiltonian\footnote{We use quotation marks because $\Psi$ still contains $k=\hbar/(2\pi)$.}:
\begin{equation}
 H_\mathrm{cl}(p,q)=T(p)+U(q)+2\Psi(p+q).
\end{equation}
The function $\Psi$ has an asymptotic behavior similar to that of $T$ and $U$:
\begin{equation}
 \Psi(Q)=\frac{\alpha}{2}|Q|+O(e^{-c|Q|}), \qquad |Q| \gg 1, 
\end{equation} 
where
\be
\alpha={N_f\over k}.
\ee
\FIGURE{
\includegraphics[height=6cm]{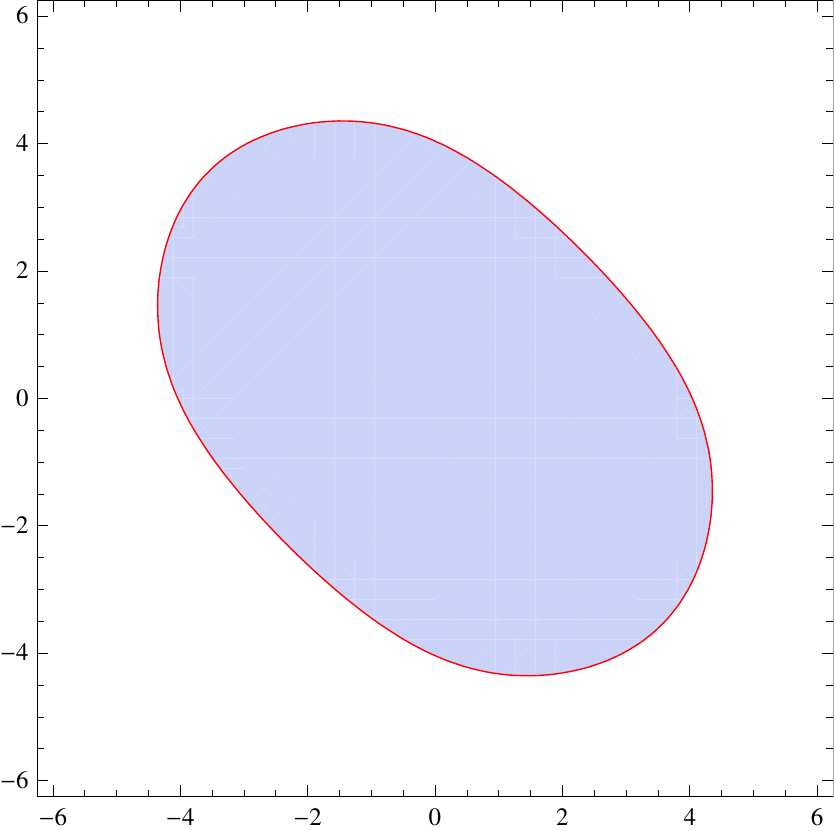} \qquad 
\includegraphics[height=6cm]{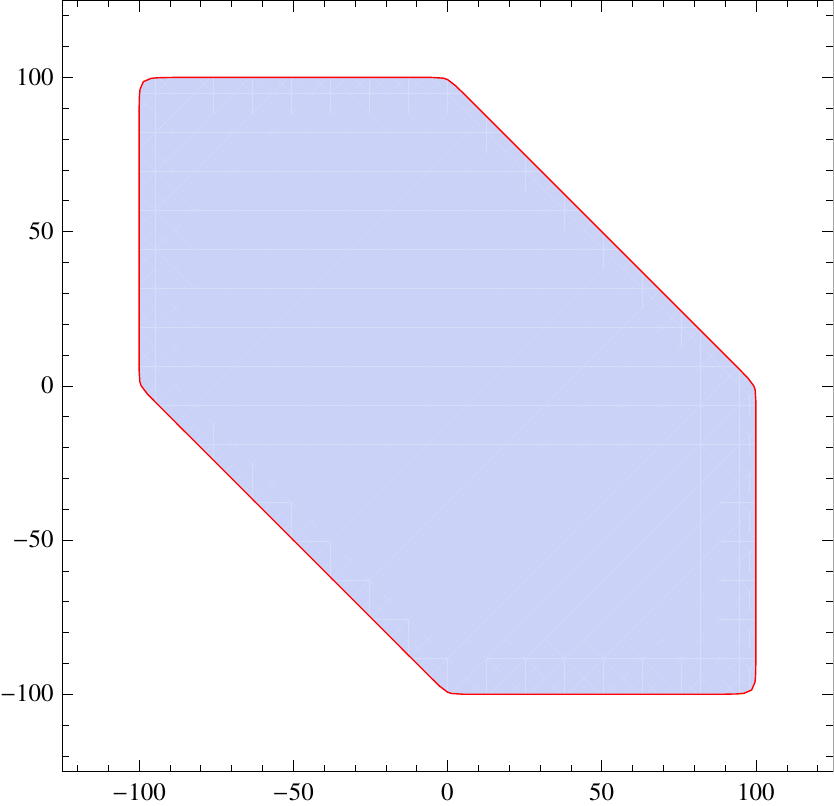}
\caption{The Fermi surface (\ref{fermi-curve}) in the $q$-$p$ plane for the ABJM theory with fundamental matter and $N_f=1$, $k=2$, for $E=5$ (left) and $E=100$ (right).}
\label{matter_fermisurface}
} 
Since $H_\wigner$ is real, (\ref{matter_KW}) is a particular example of the general case (\ref{airy_KW}) considered in the previous subsection. We have three 
different local coordinates 
\be
\label{Qsmatter}
Q_1=q, \qquad Q_2=p, \qquad Q_3=p+q.
\ee
For large energy the Fermi surface $H_\wigner(q,p)=E$ approaches a polygon given by
\begin{equation}
 \frac{|p|}{2}+\frac{|q|}{2}+\alpha|p+q|=E,
\end{equation}
see Fig.~\ref{matter_fermisurface}. Therefore, the leading contribution to the number of states is
\begin{equation}
 n(E)\approx {1 \over 2\pi\hbar} \Vol\left\{\frac{|p|}{2}+\frac{|q|}{2}+\alpha|p+q|<E\right\}=CE^2
\end{equation} 
where
\begin{equation}
 C=\frac{2(1+\alpha)}{\pi^2 k(1+2\alpha)^2}.
\end{equation}
It follows that
\begin{equation}
 F(N)\approx -\frac{\sqrt{2}}{3}\,\pi k\,\frac{1+2\alpha}{\sqrt{1+\alpha}}\,N^{3/2}
\end{equation}
which reproduces the result of \cite{cmp,hklebanov}. Notice that, as in the case of ABJM theory, the large energy limit of the Fermi surface is closely related to the 
tropical limit of the spectral curve obtained in \cite{cmp}. 

\begin{figure}
\begin{center}
\includegraphics[height=6.5cm]{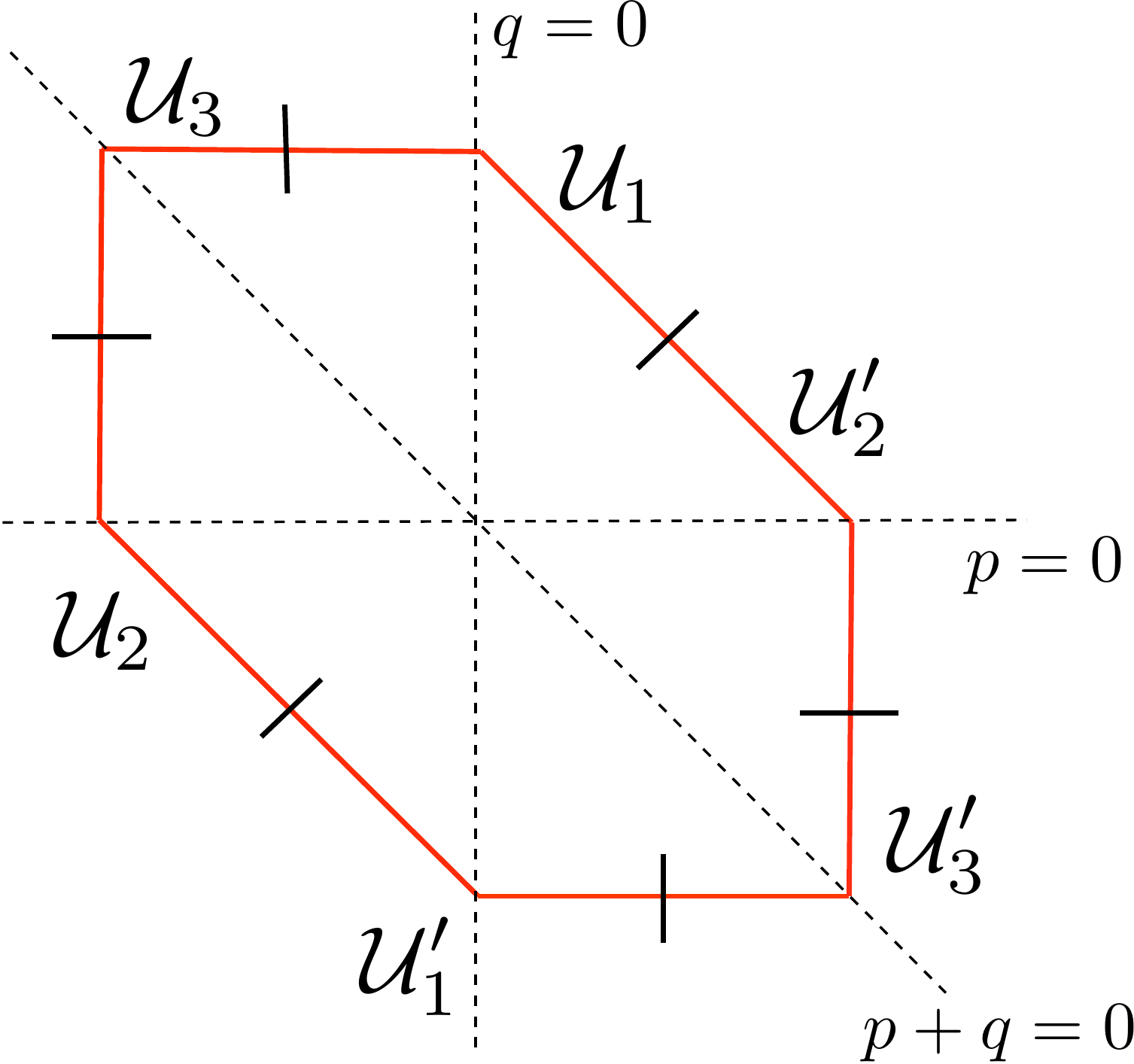}
\end{center}
\caption{The regions $\CU_R$, $\CU'_R$ defined in section \ref{airy-general} for the Fermi surface of ABJM theory with matter. The dashed lines 
are defined by $Q_R=0$, where the different coordinates are given in (\ref{Qsmatter}).}
\label{matter_regions}
\end{figure}

Now let us compute the corrections according to the general scheme described in the previous subsection. The regions 
$\CU_R, \CU'_R$ for $R=1,2,3$, as well as the lines $Q_R=0$, are shown in \figref{matter_regions}. 
In the domain $\mathcal{U}_1$ the Hamiltonian can be written as
\begin{equation}
H_\wigner(q,p)= p \left(\alpha +\frac{1}{2}\right)+U(q)+q \alpha -\frac{1}{48} \hbar ^2 (2 \alpha +1)^2 U''(q)
+\sum_{n>1}\hbar^{2n}c_n^1U^{(2n)}(q)+\CO(\re^{-cE}).
\end{equation}
Therefore we can take
\be
P_1=p+\frac{\alpha}{1/2+\alpha}q
\ee
and
\begin{equation}
 \Delta_\rp P_1(Q_1)=-\frac{2}{1+2\alpha}\left\{
U(Q_1)-\frac{|Q_1|}{2} -\frac{1}{48} \hbar ^2 (2 \alpha +1)^2 U''(Q_1)
+\sum_{n>1}\hbar^{2n}c_n^1U^{(2n)}(Q_1)
\right\},
\end{equation} 
\begin{equation}
 \Delta_\rp\Vol_1=-\frac{\pi^2}{3(1+2\alpha)}+\frac{\hbar^2 (1+2\alpha)}{24}.
\end{equation} 
In the domain $\mathcal{U}_2$ we have, 
\begin{equation}
H_\wigner(q,p)= T(p)-p \alpha -q \left(\alpha +\frac{1}{2}\right)+\frac{1}{96} (2 \alpha +1)^2 \hbar ^2 T''(p)
+\sum_{n>1}\hbar^{2n}c_n^2T^{(2n)}(p)+\CO(\re^{-cE}),
\end{equation}
therefore 
\be
P_2=-q-\frac{\alpha}{1/2+\alpha}p
\ee
and
\begin{equation}
 \Delta_\rp P_2(Q_2)=-\frac{2}{1+2\alpha}\left\{
T(Q_2)-\frac{|Q_2|}{2} +\frac{1}{96} \hbar ^2 (2 \alpha +1)^2 T''(Q_2)
+\sum_{n>1}\hbar^{2n}c_n^2T^{(2n)}(Q_2)
\right\},
\end{equation} 
\begin{equation}
 \Delta_\rp\Vol_2=-\frac{\pi^2}{3(1+2\alpha)}-\frac{\hbar^2 (1+2\alpha)}{48}.
\end{equation}
In the domain $\mathcal{U}_3$,
\begin{equation}
 H_\wigner(q,p)=p/2-q/2+2 \Psi \left(p+q\right)+\frac{1}{48} \hbar ^2 \Psi ''\left(p+q\right)
 +\sum_{n>1}\hbar^{2n}c_n^3\Psi^{(2n)}(p+q)+\CO(\re^{-cE}).
\end{equation}
Therefore 
\be
P_3={p-q \over 2}
\ee
and
\begin{equation}
 \Delta_\rp P_3(Q_3)=-\left(2\Psi(Q_3)-\alpha|Q_3|\right)
-\frac{1}{48} \hbar ^2 \Psi ''\left(Q_3\right)
-\sum_{n>1}\hbar^{2n}c_n^3\Psi^{(2n)}(Q_3),
\end{equation} 
\begin{equation}
 \Delta_\rp\Vol_3=-\frac{5}{48} \hbar ^2\alpha
\end{equation}

Finally, we obtain
\begin{equation}
  n_0=\frac{1}{\pi\hbar}\sum_{R\in \{1,2,3\}}\Delta_\rp\Vol_R=-\frac{1}{3k(1+2\alpha)}+\frac{k}{24}\left(1-3\alpha\right),
\end{equation} 
and we conclude that the partition function is given by the Airy function 
\begin{equation}
 Z(N)=C^{-1/3}\re^{A}\Ai\left[C^{-1/3}(N-B)\right]+Z_\np(N)
\end{equation} 
with 
\begin{equation}
\label{shiftmatter}
 B=n_0+\frac{C\pi^2}{3}=\frac{1}{3k(1+2\alpha)^2}+\frac{k}{24}\left(1-3 \alpha \right).
\end{equation}

The above procedure can be repeated for other quivers with a Hermitian Hamiltonian in order to determine the precise value of $B$. For example, for 
the four-node quiver with levels
\be
(k, -2k, 2k, -k),
\ee
one finds
\be
\label{shift-4}
B=-\frac{13}{135 k}+\frac{k}{8}.
\ee
Clearly, it would be nice to have a close answer for the shift for a more general class of quivers (like for example four node quivers with a Hermitian density matrix). 
In addition, it would be interesting to compare the shifts (\ref{shiftmatter}), (\ref{shift-4}) with a direct calculation from the M-theory/type IIA geometry, as in \cite{bh,ahho}.

\subsection{The massive theory}
The techniques developed in this paper can be also applied to a variant of ABJM theory in which the Chern--Simons levels $k_1$, $k_2$ do not add up to zero \cite{gt}. We will denote 
\be
2 \pi  \ri \theta=-{1\over k_1} -{1\over k_2}, \qquad {4 \pi \over \hbar}={1\over k_1} -{1\over k_2}, 
\ee
so that the original ABJM theory is recovered when $\theta=0$. Notice that $\theta$ is in principle imaginary, but it will be useful to Wick-rotate it to real 
values (see also \cite{suyamatec}). The theory with $k_1+k_2 \not= 0$ was 
studied in \cite{gt}, where it was argued that a non-zero $\theta$ corresponds to a non-zero Romans mass in type IIA 
supergravity. For this reason, we will call this theory the ``massive" theory. The massive theory was further 
investigated in \cite{ajtz}, where it was found that its free energy scales with $N$ 
as
\be
\label{53sc}
F(N)\approx \left(k_1+ k_2\right)^{1/3} N^{5/3}.
\ee
This scaling was reproduced in \cite{jkps} from an analysis of the matrix model representing the partition function, 
\be
\ba
&Z(N, \theta)\\
&={1\over N!^2} \int {\rd ^N x \over (2\pi)^N} {\rd ^N y \over (2\pi)^N} {\prod_{i<j} \left[ 2 \sinh \left( {\mu_i -\mu_j \over 2} \right)\right]^2
  \left[ 2 \sinh \left( {\nu_i -\nu_j \over 2} \right)\right]^2 \over \prod_{i,j} \left[ 2 \cosh \left( {\mu_i -\nu_j \over 2} \right)\right]^2 } 
  \exp \left[ {\ri \over 4 \pi} \sum_{i=1}^N \left(k_1 \mu_i^2 + k_2\nu_i^2 \right) \right].
  \ea
  \ee
  The exact planar resolvent of this theory was found, in a somewhat implicit form, in \cite{suyamagt}. The scaling (\ref{53sc}) can be also derived from this resolvent by 
  using the techniques of \cite{suyamatec}. 
  
In order to apply the Fermi gas picture to this theory, we have to find an appropriate density matrix. An elementary computation leads to
\be
Z(N,\xi)={1 \over N!} \sum_{\sigma  \in S_N} (-1)^{\epsilon(\sigma)}  \int  \rd ^N x \prod_i \rho (x_i, x_{\sigma(i)};\theta).
\ee
where
\be
\label{dens-xi}
\rho(x_1, x_2;\theta)={\rm e}^{-{1\over 2} U_\theta (x_1)}K(x_1, x_2;\theta)  {\rm e}^{-{1\over 2} U_\theta (x_2)}. 
\ee
Here, the one-body potential is given by
\be
U_\theta(q)=\log\left(2 \cosh {q\over 2}\right)+{\theta \over 2} q^2, 
\ee
while the function $K$ is given by 
\be
\ba
\label{Kxi}
&K(x_1, x_2;\theta)\\
&={\sqrt{1 +\hbar^2 \theta^2/4}} \int_{-\infty}^{\infty} {\rd y \over 4 \pi \hbar  \cosh {y \over 2} } \exp\left\{-{\theta\over 2} y^2  - y \left[ {\theta\over 2}(x_1+x_2) + {\ri \over \hbar} (x_1-x_2)\right]\right\}.
\ea
\ee
Although (\ref{Kxi}) is complicated, its Wigner transform is very simple, 
\be
K_{\rm W}(q,p;\theta)= {\sqrt{1 +\hbar^2 \theta^2/4}} \, \, \re^{-T_\theta(q,p)}
\ee
where
\be
T_\theta(q,p)= \log\left(2 \cosh {p\over 2}\right)+{\theta \over 2} p^2+ \theta p q, 
\ee
and of course
\be
K_W(q,p;0)=\re^{-T(p)}. 
\ee
The Wigner transform of the density matrix is then 
\be
\rho_{\rm W}(\theta)={\sqrt{1 +\hbar^2 \theta^2/4}}\, \,  {\rm e}^{-{1\over 2} U_\theta (q)} \star \re^{-T_\theta(q,p)}  \star  {\rm e}^{-{1\over 2} U_\theta (q)},
\ee
and defines the Hamiltonian of the theory through 
\be
\rho_{\rm W}(\theta)=\re_\star^{-H_{\rm W}(\theta)}. 
\ee
For $\theta=0$ we recover the density matrix of ABJM theory (\ref{rhowex}). 
\FIGURE{
\includegraphics[height=6cm]{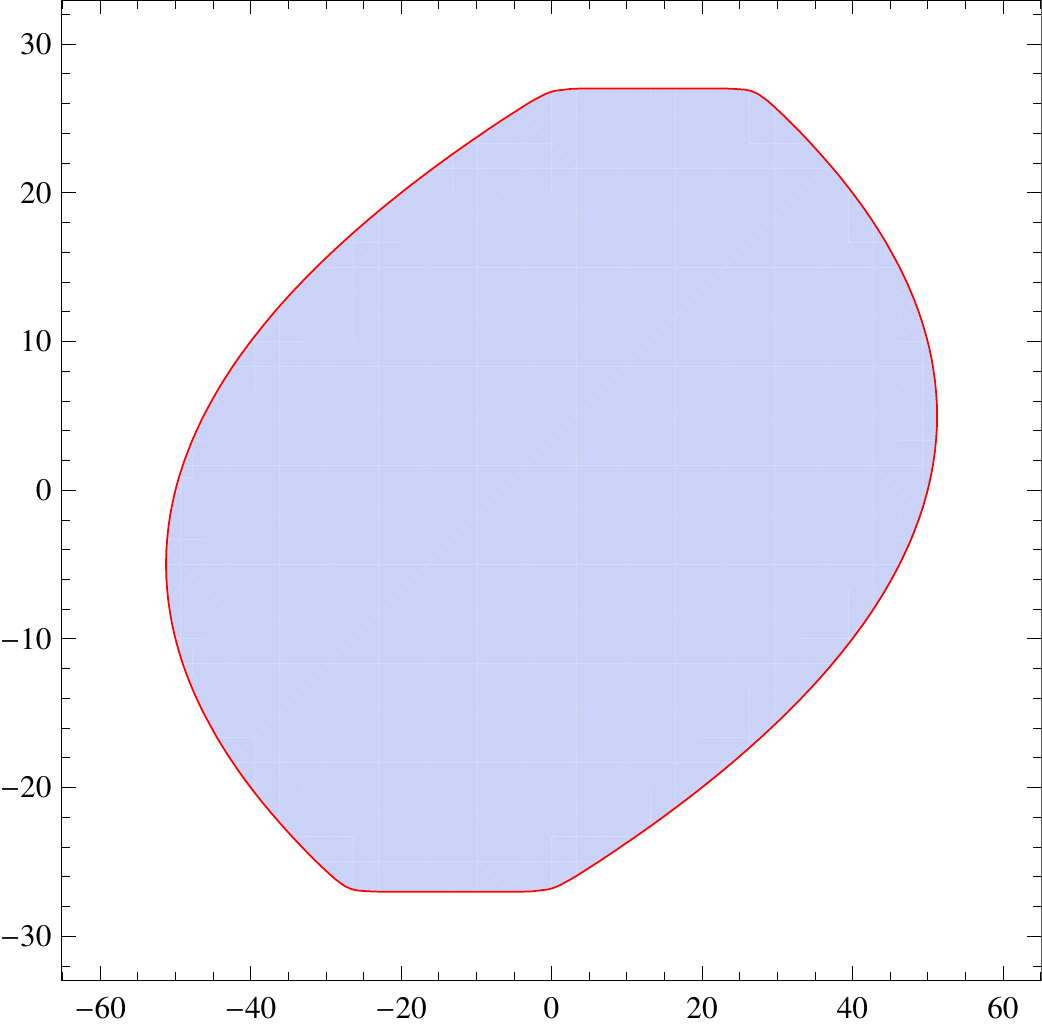} \qquad 
\includegraphics[height=6cm]{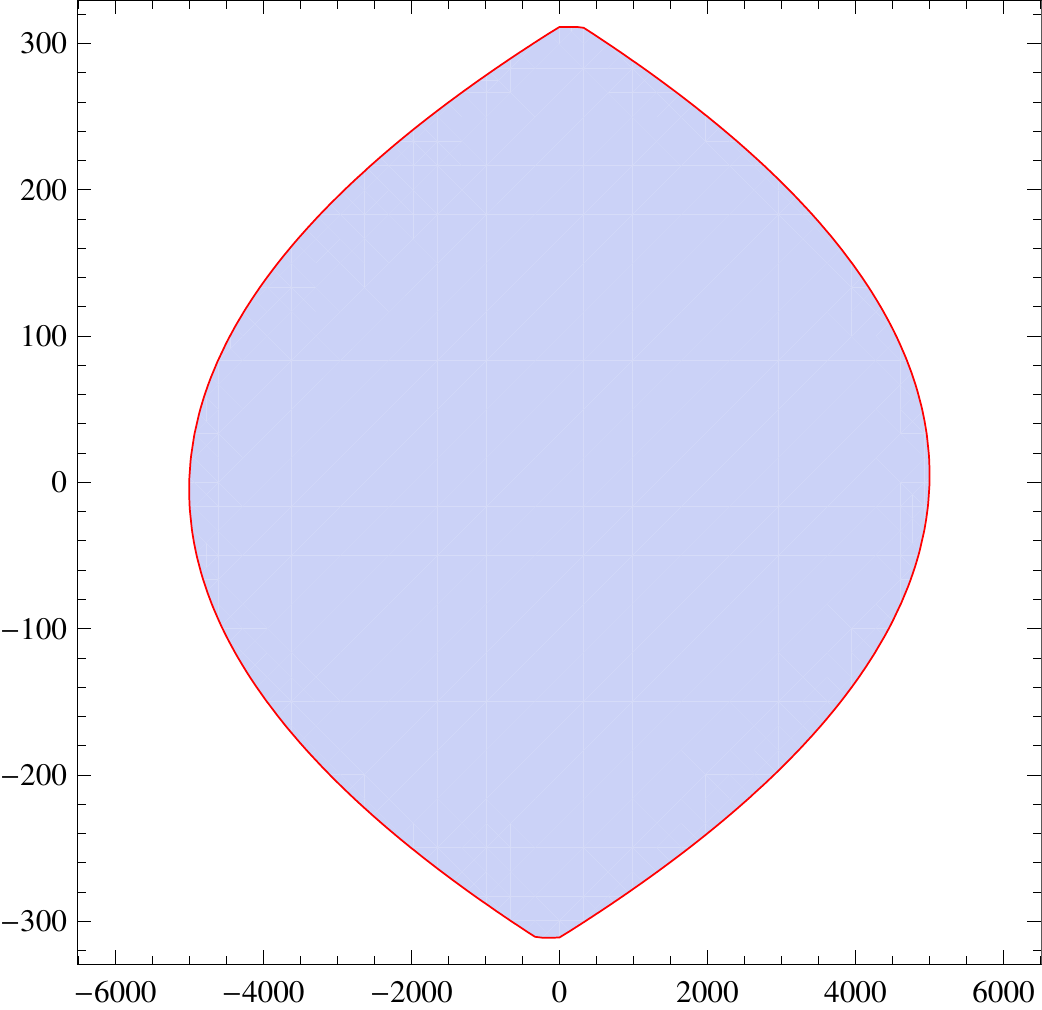}
\caption{The Fermi surface (\ref{fermixi}) in the $q$-$P$ plane, for $\theta=1/10$, $k=1$, $E=50$ (left) and $E=5000$ (right). When the energy is large, the Fermi surface approaches the surface defined by (\ref{polyxi}).}
\label{FermiMass}
}

We can now use the technology developed before to analyze the theory. We will content ourselves with an analysis of the thermodynamic limit, which 
leads to a nice interpretation of the $N^{5/3}$ behavior found in \cite{ajtz,jkps}. We will also assume that 
\be
\label{thetareg}
\left| \theta \hbar \right| \ll 1, 
\ee
or equivalently, that
\be
\left| {k_1 +k_2 \over k_1-k_2} \right| \ll 1. 
\ee
In this limit we can safely ignore quantum corrections and just look at the classical Hamiltonian 
\be
\label{Hxi}
H_{\rm cl}(q,P;\theta)=U(q) + T\left( P-q\right) +{\theta \over 2} P^2-{1\over 2} \log \left( 1+{\theta^2  \hbar^2 \over 4} \right), 
\ee
where
\be
P=p+ q. 
\ee
This linear change of variables preserves the volume form in phase space, 
\be
\rd q \wedge \rd P =\rd q \wedge \rd p. 
\ee
At large $E$ the Fermi surface
\be
\label{fermixi}
H_{\rm cl}(q,P;\theta)=E
\ee
becomes simply
\be
\label{polyxi}
{\theta \over 2} P^2 + |q|=E, 
\ee
as we can see in \figref{FermiMass}. Notice that, once $\theta\not=0$, the equation defining the Fermi surface at large $E$ has a quadratic term in the new momentum coordinate $P$ which dominates at large $E$. In other words, the Fermi gas has now a non-relativistic dispersion relation, and this changes the scaling of the free energy. 
Looking at (\ref{genH}) we deduce that 
\be
s={3\over 2}, 
\ee
therefore the free energy should scale now as $N^{5/3}$, as found in \cite{ajtz,jkps}\footnote{The matrix model analyzed in \cite{kkn} displays the same scaling.}. We find
\be
n(E) \approx \frac{4}{2\pi\hbar}\int_0^{\sqrt{\frac{2E}{\theta}}} \rd P \, 
\left(E-\frac{\theta}{2}P^2\right)=\frac{4}{3\pi\hbar}\sqrt\frac{2}{\theta}E^{3/2}.
\end{equation} 
The free energy can now be computed from (\ref{freeenergy}) and reads, 
\be
F(N) \approx -{3 \over 5 } \left( {3 {\sqrt{2}} \pi \hbar \over 8} \right)^{2/3} \theta^{1/3}N^{5/3}.
\ee
If we express this in terms of $k_1 +k_2$ we find, 
\be
\label{lead53}
F(N) \approx -{3^{5/3} \over 5 \cdot 2^{4/3}} \pi  \re^{-{\ri \pi \over 6}} (k_1 + k_2)^{1/3}  \left(1+{\theta^2  \hbar^2 \over 4} \right)^{1/3}N^{5/3}.
\ee
Since we are assuming (\ref{thetareg}), our result can be written as  
\be
F(N) \approx -{3^{5/3} \over 5 \cdot 2^{4/3}} \pi  \re^{-{\ri \pi \over 6}} (k_1 + k_2)^{1/3} N^{5/3}, 
\ee
which is precisely what \cite{jkps} obtained. Notice that, in \cite{jkps}, this result was derived based on an assumption on the behavior of the eigenvalues of the matrix model at large $N$, while here we have obtained it directly. 
When the parameter $\theta^2 \hbar^2$ is not small, one has to take into account the quantum corrections to the 
Hamiltonian, and the equation of the Fermi surface is modified. It would be interesting to study in more detail the different regimes that can occur in this theory as we vary the coupling 
constants.

\sectiono{Conclusions and prospects for future work}

In this paper we have developed an ideal Fermi gas approach to a large class of CSM theories with $\CN\ge 3$ supersymmetry. Since the particles 
in the gas do not interact, all the information of the problem is encoded in the one-particle quantum Hamiltonian. We have seen that the structure of the CSM theory determines 
the detailed form of this Hamiltonian, which is conveniently encoded in its Wigner transform $H_{\rm W}$, 
\be
\text{CSM theory} \rightarrow \text{quantum Hamiltonian} \, \, H_{\rm W}.
\ee
The semiclassical analysis of this Fermi gas already gives us a lot of information on the original partition function. As we have seen, the leading 
semiclassical contribution provides an extremely simple derivation of the $N^{3/2}$ behavior in these theories (including the right coefficient), and the next-to-leading 
correction already encodes the full $1/N$ expansion of the original matrix model. The large energy limit of the Fermi surface is a polytope which encodes the geometry of the dual tri-Sasakian target geometry. In addition, non-perturbative membrane instantons can be computed order by order 
in $k$ (but at all orders in the membrane winding). We summarize some of the features of the Fermi gas approach, as compared to the 't Hooft expansion, in the following 
table:

\begin{center}
\begin{tabular}{||  l || l || l||}
\hline
& 't Hooft expansion  & Fermi gas \\ \hline\hline
semiclassical limit& spectral curve &  Fermi surface\\ \hline
quantum corrections& quantum spectral curve & quantum Hamiltonian  \\ \hline
exponentially small 
corrections & worldsheet instantons & membrane instantons  \\\hline
non-perturbative effects & membrane instantons  & worldsheet instantons  \\\hline
\end{tabular}
\end{center}

It would be interesting to understand at a deeper level the Fermi gas picture. Since D-branes behave as fermions (see for example \cite{tsih}), and the gauge theories 
we have considered have D-brane realizations, one might be able to derive this picture directly from the D-branes underlying the gauge theory. 

Our formalism has many similarities with recent developments matrix models and topological string theory 
(see, for some examples, \cite{tsih,eo,cheng}). The semiclassical limit 
gives a Fermi surface, which can be regarded as the counterpart to the spectral curve in the conventional 't Hooft expansion. For ABJM theory both curves 
are the same, but for the more general CSM theories studied in this paper, the semiclassical Fermi surface is simpler and easier to find. 
In fact, the planar resolvent of the CSM matrix models is in general not known, and when 
it is known (as in \cite{suyamagt,cmp}), it turns out to be a complicated transcendental function. It seems however that both, the Fermi surface and the spectral 
curve, have the same tropical or polygonal limit. 

The Fermi surface has quantum corrections, and one then obtains a ``quantum corrected curve" 
nicely encoded in the function $H_{\rm W}$ in phase space. The quantum corrections appear in a very natural way, by replacing standard products 
by $\star$ products. However, the corrections found in this way are not in the string coupling constant $g_s$, as in conventional 
topological string theory \cite{tsih,gs} but rather in the inverse string coupling 
$1/g_s$. Therefore, the quantum Hamiltonian $H_{\rm W}$ is computing the quantum curve for the {\it strongly coupled} string. 

More generally, our formalism seems appropriate to study the strong-coupling regime of matrix models and topological strings. 
For example, in the case of ABJM theory, it is known \cite{mp,dmp} that the ABJM matrix model (\ref{abjmmatrix})
corresponds to the submanifold $T_1=T_2$ of the moduli space
of topological strings on local $\IP^1 \times \IP^1$. Here, $T_{1,2}$ are the K\"ahler parameters measuring the sizes of the two $\IP^1$s in the geometry. 
However, by looking at for example (\ref{topJ}), it is clear that the grand canonical potential of ABJM theory seems to be directly related to the 
topological string free energy in the large radius frame. Therefore, our calculation of $J(\mu)$ for the ABJM theory can be interpreted as a concrete strong coupling expansion of this topological string free energy, including non-perturbative effects. The worldsheet instantons of the topological string at large radius would appear then as quantum-mechanical instantons of the Fermi gas. Notice also that the grand canonical partition function, which is the focus of this paper, involves the sum over fluxes  first considered in the context of topological strings in \cite{tsih}, and studied from the matrix model point of view in \cite{eynard,em,mpp}. Our formalism gives a concrete approach to calculate this object at strong coupling, 
but one should clarify the relation between the picture proposed here and the non-perturbative approach of \cite{eynard,em} involving 
theta functions on the spectral curve. It would be very interesting to develop further all these relationships to topological string theory. 

If we stay in the world of CSM theories and their AdS duals, an obvious and important question is to which extent the Fermi gas formalism can be applied to other theories. For example, one should consider, already in the $\CN=3$ case, the issue of quantum corrections and possible Airy behavior in theories where the free energy is not 
real, as well as in theories where the nodes have different ranks \cite{abj}. The extension to $\CN=2$ theories is more challenging. In that case 
the interaction among eigenvalues involves a more complicated function \cite{jafferis,hama} and they lead generically to interacting Fermi gases, rather than to ideal gases. 

Although the equivalence between the matrix models and the Fermi gas partition function is an exact statement at finite $N$ and $k$, in this 
paper we have worked in the thermodynamic limit (large $N$) and in the semiclassical limit (expansion in powers of $k$ around $k=0$). Fortunately, as we have seen, the expansion in $1/N$ satisfies some sort of ``non-renormalization" property and we can determine it at finite $k$ by a next-to-leading computation in the WKB expansion. An obvious challenge is to solve the Fermi gas problem at finite $k$, say $k=1$, to make full contact with the M-theory expansion. This would amount to a resummation of the 
non-perturbative effects computed in this paper order by order in $k$. One possible route to achieve this is to find the exact eigenvalue 
spectrum of the quantum Hamiltonian, or equivalently, of the density matrix $\hat \rho$. In the case of ABJM theory, this means solving the integral equation (\ref{eigenv}), 
which can be written as 
\be
\int \rd x' \rho(x,x') \phi_n(x')=\lambda_n \phi_n(x), 
\ee
where the kernel can be written in the form
\be
\rho(x,x')={\re^{-{1\over 2} U(x) -{1\over 2} U('x)}\over 4 \pi k \cosh \left( {x -x' \over 2 k} \right)}. 
\ee
This type of kernel appears in other contexts, like the $O(2)$ matrix model \cite{kostov} (albeit with a different $U(x)$), and it is connected to both the Hirota hierarchy \cite{kostov,twone} and to the Thermodynamic Bethe Ansatz \cite{zamo,twtwo}. At least for $k=1,2$ (where supersymmetry is enhanced), we anticipate a nice solution to the eigenvalue problem in terms of an integrable system. 
In particular, the relation to differentiable hierarchies of the Hirota type suggests that the 
conjecture \ref{airycon} might be proved by performing a suitable double-scaling limit in the hierarchy.

It would be also very interesting to develop further 
the relationship between worldsheet instantons and quantum mechanical instantons of the Fermi gas sketched in section \ref{qmisec}. 
It is clear that the solution of the ABJM theory found in \cite{dmp}, in the 't Hooft expansion, is extremely powerful in order to capture these 
corrections, but it would be important for the development of the Fermi gas approach to have a better understanding of this issue. 

Finally, there might be a connection between the Fermi gas of this paper and two other pictures for string/M-theory based on fermions: the droplet picture 
proposed in \cite{berenstein,llm} to analyze 1/2 BPS operators in $\CN=4$ SYM theory, and the Fermi liquid picture of non-critical M-theory proposed in \cite{horavak}. 

\section*{Acknowledgements}
We would like to thank Luis \'Alvarez-Gaum\'e, Thierry Giamarchi, Daniel Jafferis, Volodya Kazakov, Igor Klebanov, 
Corinna Kollath, Ivan Kostov, Juan Maldacena, Carlos N\'u\~nez and Cumrun Vafa for useful 
conversations and correspondence. We are particularly grateful to Nadav Drukker for a detailed reading of the manuscript. 
M.M. is grateful to the Simons Center for Geometry and Physics, 
and specially to Martin Rocek, for hospitality in the first stages of this project. This work is supported by the Fonds National Suisse, subsidies 200020-126817 and 
200020-137523. P.P. is also supported by FASI RF 14.740.11.0347.

\appendix 

\sectiono{Quantum corrections in ABJM theory at order $\CO(\hbar^4)$}
In this Appendix, we give some details on the computation of the order $\CO(\hbar^4)$ corrections to the grand canonical potential of ABJM theory, 
which confirm the general arguments of section \ref{qcorr}. 

The Baker--Campbell--Hausdorff formula applied to (\ref{rhowex}) gives
 \be
 \label{Hww}
 \ba
 H_{\rm W}(q,p)&=T+U+\frac{1}{12}\left[T,\left[T,U\right]_\star \right]_\star
 +\frac{1}{24}\left[U,\left[T ,U \right]_\star \right]_\star+ {1\over 360} [[[[T,U]_\star,U]_\star, U]_\star, T]_\star\\
 &-{1\over 480}[[[[U,T]_\star,U]_\star, T]_\star, U]_\star +{1\over 360}[[[[U,T]_\star,T]_\star, T]_\star, U]_\star +{1\over 120}[[[[T,U]_\star,T]_\star, U]_\star, T]_\star\\
 &+{7\over 5760} [[[[T,U]_\star,U]_\star, U]_\star, U]_\star -{1\over 720} [[[[U,T]_\star,T]_\star, T]_\star, T]_\star +\cdots
\ea
\end{equation}
This leads to the next correction to the Wigner transform of the Hamiltonian
\be
\label{Hwtwo}
\ba
H_{\rm W}^{(2)}&={1\over 144} T'(p) T'''(p) U^{(4)}(q) -{1\over 288} U'(q) U'''(q) T^{(4)}(q) 
\\&-{1\over 240} \left(U'(q)\right)^2 U''(q) \left(T''(p)\right)^2+{1\over 60}\left(T'(p)\right)^2 T''(p) \left(U''(q)\right)^2 \\
&-{1\over 80} \left(U'(q)\right)^2 U''(q) T'(p)T'''(p) +{1\over 120}  \left(T'(p)\right)^2 T''(p) U'(q)U'''(q)\\
&+{7 \over 5760} \left(U'(q)\right)^4 T^{(4)}(p) -{1\over 720}\left(T'(p)\right)^4 U^{(4)}(q). 
\ea
\ee
The computation of $J_2(\mu)$ also involves the $\CG_r$ defined in (\ref{Gr}) up to order $\CO(\hbar^4)$. Due to (\ref{GrEx}) only the terms with 
$r\le 6$ are needed. A long but straightforward calculation leads finally to 
\be
J_2(\mu)=-{\pi^2 \over 4320}- \frac{\pi^2}{2880} \left(104+ 5 \pi^2 -134 \mu + 30 \mu^2 \right) \re^{-2\mu} +\CO\left(\mu^2 \re^{-4 \mu}\right). 
\ee
Notice that no polynomial in $\mu$ is generated, as expected from the analysis in section \ref{qcorr}.

\end{document}